\documentclass[journal,final,twocolumn]{IEEEtran}
\usepackage{comment}

\usepackage[pdftex]{graphicx}
\graphicspath{{./figures/}} 
\DeclareGraphicsExtensions{.pdf} 

\usepackage[cmex10]{amsmath}
\usepackage{algpseudocode}
\usepackage{enumitem}

\usepackage{hyperref}  

\ifCLASSOPTIONcompsoc
  \usepackage[caption=false,font=normalsize,labelfont=sf,textfont=sf]{subfig}
\else
  \usepackage[caption=false,font=footnotesize]{subfig}
\fi

\usepackage{xcolor}

\usepackage{booktabs}
\usepackage{multirow}
\usepackage{arydshln}

\begin{document}

%
\title{Fast 2D Convolutions and Cross-Correlations Using Scalable Architectures}
%
%
%

\author{Cesar~Carranza,~\IEEEmembership{Student Member,~IEEE,}
        Daniel~Llamocca,~\IEEEmembership{Senior Member,~IEEE,} \\
        and~Marios~Pattichis,~\IEEEmembership{Senior Member,~IEEE}
\thanks{C. Carranza is with Secci\'on Electricidad y Electr\'onica, 
         Pontificia Universidad Cat\'olica del Per\'u, Lima-32, Per\'u 
         (e-mail: acarran@pucp.edu.pe).
    He has been with the Department of Electrical and Computer Engineering, 
         University of New Mexico, Albuquerque, NM, 87131 USA. 
   }
\thanks{M. Pattichis is with the Department
         of Electrical and Computer Engineering, University of New Mexico, 
         Albuquerque, NM, 87131 USA (e-mail: pattichis@ece.unm.edu)}
\thanks{D. Llamocca is with the Electrical and Computer Engineering Department, Oakland University, Rochester, MI, 48309, USA (e-mail: llamocca@oakland.edu)}

\thanks{}
}

\markboth{IEEE Transactions on Image Processing,~Vol.~XXX, No.~X, XXXX~XXXX}%
{Carranza \MakeLowercase{\textit{et al.}}: Fast 2D convolutions Using Scalable Architectures}

\maketitle

\begin{abstract}
The manuscript describes fast
  and scalable architectures and associated algorithms for computing
  convolutions and cross-correlations. 
The basic idea is to
   map 2D convolutions and cross-correlations to
   a collection of 1D convolutions and cross-correlations
   in the transform domain.
This is accomplished through the use
   of the Discrete Periodic Radon Transform (DPRT)
   for general kernels and the
   use of SVD-LU decompositions for low-rank kernels.

The approach uses scalable architectures
   that can be fitted into modern FPGA and Zynq-SOC devices.
Based on different types of available resources, for $P\times P$ blocks, 
   2D convolutions and cross-correlations 
   can be computed in just $O(P)$ clock cycles
   up to $O(P^2)$ clock cycles.
Thus, there is a trade-off between
   performance and required numbers and types of resources.
We provide implementations
   of the proposed architectures using
   modern programmable devices (Virtex-7 and Zynq-SOC).
Based on the amounts and types of required resources,
   we show that the proposed approaches 
   significantly outperform current methods.
\end{abstract}

\begin{IEEEkeywords}
Linear Convolution, Circular Convolution, Cross-correlation, Discrete Periodic Radon Transform, Parallel Architecture, Scalable Architecture, FPGA, SOC.
\end{IEEEkeywords}

\IEEEpeerreviewmaketitle

\section{Introduction} \label{sec:intro}
\IEEEPARstart{C}{onvolutions} and cross-correlations have wide applications
    in image and video processing and imaging \cite{bovik2009a,bovik2009b}.
The development of effective architectures and algorithms
    for computing convolutions
    and cross-correlations can be used in several applications
    (e.g., feature extraction \cite{Nixon2012}, 
           template matching \cite{Anam2012}, 
           pattern recognition \cite{Anam2014},     
           edge detection, filtering, deconvolution, segmentation, 
           and denoising 
           \cite{bovik2009a}).

To support implementations in modern devices (e.g., FPGAs, PSOCs), we
  are also interested in scalable architectures.
The basic idea is to make efficient use of hardware resources to deliver
  the best possible performance.
For scalability, we investigate implementations
   that can be fitted within available resources.

A standard approach for developing efficient architectures
   for 2D convolutions and cross-correlations would be to build the systems based on 2D FFTs.
As is well-known (e.g., see \cite{Dudgeon1990,Rao2010}), 
   for sufficiently large
   kernels, the use of 2D FFTs will give better results than
   direct approaches.
Unfortunately, the direct implementation of 2D FFTs in hardware
   requires the use of complex-valued
   arithmetic units.   
As a result, the hardware scalability of using 2D FFTs is 
   fundamentally limited
   by the number of 1D FFT processors that can be fitted in
   any given hardware device.
We refer to \cite{Uzun2005, Xilinx2012,Ayinala2012} 
   for details of the latest implementation
   of this approach.
As shown in \cite{Uzun2005}, performance can be
   improved by including up to 4 1D FFT processors.
Beyond 4 1D FFT processors, performance stalls or even degrades
   due to I/O issues \cite{Uzun2005}.

Modern FPGA and SOC devices are equipped with DSPs that can better facilitate the implementation
      of 2D FFTs.
   Thus, in modern FPGAs and SOCs, 
     the scalability of the FFT-based methods is largely limited by the number of
     available DSPs.
To provide fair comparisons, for our FPGA and SOC implementations, 
     we compare our proposed approaches against the use of DSPs in FFT-based methods. 

The development of $O(P)$ methods for 2D convolutions also poses significant I/O issues.
  For example, a sequential access through the pixels will require 
     $O(P^2)$ clock cycles.
  In our proposed designs, we use parallel loads and stores that
     can move entire rows and columns of blocks of $P$-pixels into an
     array of registers in a single clock cycle.
  Furthermore, we rely on the use of parallelism and 
     pipelined designs to ensure that
     each row of pixels is also processed in $O(P)$ clock cycles.
     
The number of clock cycles to compute the 1D FFT
    can be reduced from $O(P \log_2 P)$ to 
    $O(P)$ using a fully-pipelined
    and parallel hardware implementation
    as documented in \cite{Ayinala2012}.
To provide fair comparisons, we 
   will also consider 2D extensions of this work.
Overall, due to the need to implement complex
   arithmetic operations for this approach, we have found that
   actual hardware resources remain relatively high
   for many reasonable values of $P$.   

Alternatively, two-dimensional convolutions and cross-correlations can also be computed
  in the transform domain using the
  2D Discrete Periodic Radon Transform (DPRT).
The DPRT can be computed using summations
  along different directions \cite{carranza2014,carranza2015}.
Similar to the FFT, the DPRT approach requires
  that we first take the DPRT of the image and the 2D kernel.
Then, along each DPRT direction, we compute
  1D circular convolutions/cross-correlations between the
  DPRTs of the image and the 2D kernel.
The 2D convolution/cross-correlation result can then be computed
  by taking the inverse DPRT of the previous result.
Unlike the 2D FFT approach, the DPRT can be implemented
  with real-valued fixed-point additions. 
Furthermore, as shown in \cite{carranza2015},
  we now have fast and scalable fixed-point
  architectures that can be implemented in FPGAs or SOCs
  that can compute DPRTs in $O(P)$ to $O(P^2)$ clock cycles
  depending on available hardware resources.

To implement fast and scalable convolutions and cross-correlations
  based on the DPRT, we also need to compute the separable 1D convolutions/cross-correlations
  in $O(P)$ and $O(P^2)$ clock cycles and address I/O issues
  when connecting with the DPRT blocks.
For the fastest approach, we 
  develop 
  \textit{FastConv} and \textit{FastXCorr}
  based on the fast DPRT that
  can compute convolutions/cross-correlations in $O(P)$ clock cycles.
Similarly, we develop 
  \textit{FastScaleConv} and \textit{FastScaleXCorr} 
  based on the scalable DPRT.   

To provide balanced comparisons to other approaches,
   we also discuss resource requirements.
Here, we restrict our discussion
   to the numbers of required 
   additions, multipliers, and flip-flops.
In general, we will classify
   and approach as one of requiring
   $O(P^2)$ or $O(P^3)$ resources
   if the number of additions, multipliers, or flip-flops
   grows as $O(P^2)$ or $O(P^3)$ respectively.
We will provide more detailed comparisons
   in terms of the exact numbers
   of additions, multipliers, or flip-flops, DSPs, and other
   types of resources
   in a later section of the paper. 

We also define scalability based on available resources.
  We are interested in convolution systems that can
     be scaled so as to fit into different device sizes.
  Then, the fastest methods refer to approaches that can compute
     2D convolutions using the minimum number of clock cycles
     while requiring the maximum amount of resources.
  On the other hand, slower implementations will
     require fewer resources.
Thus, we have a clear trade-off between performance
     and required resources.
		
In addition to comparisons against FFT based methods,
   we also consider spatial-domain methods based on 
   systolic arrays \cite{Kung1982}.
The standard systolic array implementation of 1D convolutions
   computes an output every clock cycle.
Without using separability, a direct extension of the 1D systolic
   array approach requires that we keep several image rows in memory
    \cite{Hon1990,Mohanty1996}.
As a result, the application of non-separable systolic array
   implementations has been limited to relatively small kernels.
Furthermore, a derivation of a $O(P)$ clock cycles approach
   based on systolic approaches leads
   to prohibitive hardware resource growth of $O(P^3)$ \cite{Hon1990}.
In comparison, hardware resourses in all of our proposed methods
   only grow as $O(P^2)$ or less.

In the spatial domain, we also have the relatively recent emergence
    of fast convolution using a sliding window \cite{dong2007optimized}.
At each image pixel, a sliding window of the same size as the kernel
   is applied to compute one output pixel
   of the convolved image \cite{fowers2012performance}.
This comes at a cost of using as many multipliers and adders as 
   the coefficients in the kernel,
   and thus grows linearly with the number of coefficients in the kernel.

We also develop 
   \textit{FastRankConv},
   a second family of fast and
   scalable architectures that represents an extension
   of the current systolic methods.
Our approach is based on the use of 
   separable approximations of non-separable
   kernels \cite{Antoniou1990,Antoniou2005}.
The basic idea is to express non-separable kernels as
   a sum of of a small number of separable kernels.
Then, scalable hardware implementations can be derived
   by controlling the number of efficient 1D processors.

Overall,
   we describe and implement two fast methods for computing 2D convolutions in $O(P)$ 
   clock cycles.
Both methods map 2D convolutions
   into a collection of 1D convolutions that are computed
   in $O(P)$ clock cycles.
Each 1D convolution is computed in parallel
   using a row of multipliers
   followed by an adder tree.
For \textit{FastRankConv}, we approximate the 2D convolution
   kernel using a minimal number of 1D kernels that
   are applied along each row and each column.
For \textit{FastConv} and \textit{FastScaleConv},   
  we first take the DPRT in $O(P)$ clock cycles using
  the fast DPRT,
  compute the 1D convolutions in the transformed domain,
  and then take the inverse DPRT in $O(P)$ clock cycles
  using the inverse DPRT. 

We summarize the primary architectural elements
of our design:
\begin{itemize}
        \item \textit{\textbf{An array of circular-shift-registers:}}
    The image data is processed
        using an array of circular shift registers.
  
    \item \textit{\textbf{Fast memory:}}
    The memory array is implemented using a row of
        SRAMs where each SRAM stores a column of the image.
 
        \item \textit{\textbf{Row-level parallel I/O:}}
        The scalable architectures load the image into memory
           using a sequence of parallel loads of rows of pixels.
        Thus, for an image with $N$ rows,
           we can load the entire image into memory in $N$ cycles.
        
        \item \textit{\textbf{Row-level parallel and pipelined processing:}}
        The proposed scalable architectures are designed to process
           multiple rows at the same time.
        Thus, for FPGA and SOC implementations,
           the idea is to implement as many row-processing units
           as we can fit in the device.
        Then, each row-processor uses a pipelined architecture
           that produces results after each cycle after an initial latency.                         
        
        \item \textit{\textbf{Fast transpositions:}}
        We significantly reduce the transposition overhead
            using an additional output memory array.
        The output memory array uses dual-port memories
            to allow us to write the output results and read
            intermediate values at the same time.
        Based on our proposed approach, we can read and write
            rows and columns in a single cycle as needed.
        Overall, in our pipelined design, the net effect
            is that transposition is performed during computation
            and will thus not require any additional cycles.
\end{itemize}

The scalability characteristics of our proposed
        architectures include:
\begin{itemize}                 
        \item \textit{\textbf{Performance scalability by
                        controlling the number of row-processors
                        in the DPRT and the 1D convolutions/cross-correlations:}}
         We refer to \cite{carranza2015} for the scalable DPRT implementation.
        
        \item \textit{\textbf{Pareto optimality:}}
         We present Pareto-optimal designs in the sense that
            our family of architectures provide the fastest implementations
            based on available resources.
         In other words, additional resources always yield faster performance.
        
        \item \textit{\textbf{Fast 2D convolutions and
                      cross-correlations:}}
         \textit{FastConv} and \textit{FastXCorr}         
             compute convolutions and cross-correlations
             for $P\times P$ blocks in $O(P)$.    
         For large images, 
             the image can be broken into $L$ separate
             blocks of size $P \times P$ and use an overlap-and-add approach
             to compute the final results.
         Thus, in the fastest case,
             we can compute convolutions and cross-correlations
             in just $O(L\cdot P)$ clock cycles.
         On the other hand, in the worst case scenario,
             with very limited resources,
             2D convolutions and cross-correlations
             can be computed
             in $O(L\cdot P^2)$ clock cycles.
         Here, we use the term \textit{large image} to refer to image sizes
             that require more on-chip resources than what is available.                   
\end{itemize}

The rest of the manuscript is organized as follows.
The mathematical definitions for the DPRT,
   its inverse, and the transformation property of the DPRT
   are given in section \ref{sec:background}.
The proposed approach is given in section \ref{sec:methods}.
Section \ref{sec:results} presents the results. Conclusions and future
work are given in section \ref{sec:conclusions}.

\section{Background}\label{sec:background}

\subsection{Basic notation}\label{sec:notation}
Let  $g(i,j)$ denote an image (or image block) of $P_1$ rows with $P_2$ pixels per row be
    of size $P_1 \times P_2$ with $B$ bits per pixel.
We index $g(i,j)$ using  $0\leq i \leq P_1-1$ and $0\leq j \leq P_2-1$.
We use $h$ to denote the convolution kernel and assume
    a size of $Q_1 \times Q_2$ with $C$ bits per pixel.
We use $f(i,j)$ for the output of size
    $N_1 \times N_2$ where $N_1=P_1+Q_1-1$ and $N_2=P_2+Q_2-1$.
For the case when $N_1=N_2$ and $P_1=P_2$, 
    we simply use $N$ and $P$ throughout the text. 

Typically, images tend to be much larger than the convolution or cross-correlation kernels.
In such cases, the input image $g$ will be broken
   into blocks that are equal to the size of the kernel.
Thus, in the most typical scenario, 
  we assume that the image and kernel blocks are of size
  $P\times P$.
After linear convolution, the output image block is 
  of size $N\times N$ where $N=2P-1$.
To compute outputs over the entire image,
  we break the image into non-overlapping blocks
  and use overlap-and-add to produce the final results.     
  
\subsection{Separable decomposition for non-separable kernels} \label{sec:LUdecomp}
We begin with the 2D Z-transform of the convolution kernel $h$:
\begin{equation}
    H(z_1, z_2)
          = \sum\limits_{i=0}^{Q_1-1}
            \sum\limits_{j=0}^{Q_2-1}
            h(i, j) z_1^{-i} z_2^{-j}.
            \label{eq:z-transf}
\end{equation}
To allow for separable decompositions,
   we consider a matrix re-formulation of
   \eqref{eq:z-transf} \cite{Antoniou2005}:
\begin{equation}
    H (z_1, z_2) =
     \mathbf{Z_1}^T \, \mathbf{H} \, \mathbf{Z_2}
   \label{eq:h-exp}
\end{equation}
where we have placed all of the filter coefficients in
  $\mathbf{H}$, and
  $\mathbf{Z_i}= [1 \, z_i^{-1} \, z_i^{-2} \, \dots \, z_i^{-(Q_i-1)}]$
  for $i=1, \, 2$.
Now that we have the filter coefficients in matrix form,
  we can consider separable matrix approximations to $\mathbf{H}$.
First, consider the singular value decomposition (SVD)
  for $\mathbf{H}$:
  $\mathbf{H}=\mathbf{U}\mathbf{\Sigma}\mathbf{V}^T$.
Then, we can simplify $\mathbf{H}$ by zeroing
  out the smallest singular values of $\mathbf{\Sigma}$.
If we let $\mathbf{\Sigma_m}$ denote the resulting
  $\mathbf{\Sigma}$ after zeroing-out small singular values,
  we reconstruct an effective approximation to $\mathbf{H}$
  using $\mathbf{H_r} = \mathbf{U}\mathbf{\Sigma_r}\mathbf{V}^T$
  where we have kept the $r$ larges singular values of $\mathbf{H}$.
In this case, we use the LU decomposition of $\mathbf{H_r}$
   to get  \cite{Antoniou2005}:
\begin{equation}
H_r (z_1, z_2) = \sum\limits_{k=1}^{r}
  \left(\sum\limits_{i=0}^{Q_1-1} l_{ki}^m z_1^i\right)
  \left(\sum\limits_{j=0}^{Q_2-1} u_{jk} z_2^j\right)
  \label{eq:separable}
\end{equation}
where $r$ also denotes the rank of $\mathbf{H_r}$.
In \eqref{eq:separable}, we have
   expressed the original 2D convolution
   into a sum of $r$ separable
   1D convolutions along the rows and columns.
Furthermore, it is clear that the separable
   decomposition also applies to non-separable
   2D kernels.
Furthermore, in the simplest case we have
   $r=1$ which eliminates the hardware required
   for accumulating the additions.
We will not consider this case any further (see \cite{llamocca2011} for details).

\subsection{The discrete periodic radon transform (DPRT)}\label{sec:dprt-only}
We define the DPRT of $f$ of size $N \times N$, $N$ prime, using \cite {Hsung1996}:
\begin{equation}
    F(m,d) = \left\{
    \begin{array}{ll}
            \sum\limits^{N-1}_{i=0}f(i,\left\langle d+mi\right\rangle _{N}), & 0 \leq m < N, \\ 
            & \\
            \sum\limits^{N-1}_{j=0}f(d,j), & m = N, 
          \end{array}
          \right.
          \label{eq:dprt}
\end{equation}
where $d = 0, 1, \dots, N-1$,
      $m = 0, 1, \dots, N$, and
      $<.>_N$ denotes the positive remainder
      when we perform integer division by $N$ (e.g.,
      $<128>_{127}=1$).
In \eqref{eq:dprt}, we have that $m$ indexes the prime directions.
Along each prime direction, we add up the pixels along each ray.
In \eqref{eq:dprt}, $d$ is used to index each the rays of each direction.

The inverse DPRT can be used to reconstruct $f$
   from the forward DPRT using:
\begin{equation}  \label{eq:idprt}
    f\left(i,j\right) =\frac{1}{N}\left[
      \sum^{N-1}_{m=0} F\left(m,\left\langle j-mi\right\rangle_{N}\right)
                       -S+F\left(N,i\right)\right]
\end{equation}
where:
\begin{equation}  \label{eq:fullsum}
    S =\sum^{N-1}_{j=0}\sum^{N-1}_{i=0}f(i,j).
\end{equation}
As noted in the definition, the size of the transform needs to be restricted to prime numbers.
We do not impose this restriction directly to the input image block and kernel sizes, but to the result
   of the linear convolution of size $N_1 \times N_2$, with $N_1=P_1+Q_1-1$ and $N_2=P_2+Q_2-1$.
   Therefore, a minimal (or even none) zero padding is required if the input sizes are selected conveniently.
There are several reasons for imposing this restriction.
Most importantly though, for prime $N$, the DPRT provides the most efficient implementations by requiring the minimal
   number of $N+1$ primal directions \cite{kingston2006projective}.
It is important to note that prime-numbered transforms
  have advantages in convolution applications.
Here, just like for the Fast Fourier Transform (FFT),
  we can use zero-padding to extend the DPRT
  for computing convolutions in the transform domain.
Unfortunately, when using the FFT with $N=2^p$, zero-padding requires
  that we use FFTs with double the size of $N$.
In this case, it is easy to see that the use of prime-numbered DPRTs
  is better since there are typically many prime numbers between
  $2^p$ and $2^{p+1}$.

We refer to \cite{carranza2015} for fast and scalable implementations
   of the DPRT and its inverse.
In the fastest case, we can compute  the full DPRT in
   just $2N+\left\lceil \log_2N\right\rceil+1$ clock cycles with $O(N^2)$ growth
   in resource usage.
For the scalable DPRT implementation,
   we require $\left\lceil N/H\right\rceil(N+3H+3)+N+\left\lceil \log_{2}H\right\rceil+1$ cycles
   where $H$ is used as the scalability parameter.
We have a family of scalable DPRT implementation using
   $H=2,\ldots,N$
   with a resource usage that grows from 
   $O(N)$ for the slowest case ($H=2$)
   to $O(N^2)$ for the fastest case ($H=N$).

\subsection{Circular convolutions using the DPRT}

Consider the 2D circular convolution
   $f = g \otimes h$ given by:
\begin{equation}  \label{eq:conv2d}
    f(k,l)  =  \sum^{N-1}_{i=0}\sum^{N-1}_{j=0}g(i,j)h(\left\langle k-i\right\rangle_{N},\left\langle l-j\right\rangle_{N}).
\end{equation}
To define the DPRT convolution property, let $m$
   denote a prime direction and
   define the DPRTs along the $m$-direction using:
     $F_m(d)=F(m, d), G_m(d)=G(m, d), H_m(d)=H(m,d)$.
We then have that the $m$-direction DPRTs are related
   through 1D dimensional circular convolution in the
   transform domain as given by \cite{Lun1995}:
\begin{equation}
    F_m (d) = \sum^{N-1}_{k=0} G_m (k) H_m (\left\langle d- k\right\rangle_{N}) \\
               \label {eq:conv1dshort}
\end{equation}
Thus, we can compute the result of 2D circular convolution
      in the transform domain using 1D circular convolutions along
      all of the prime directions as given by \eqref{eq:conv1dshort}.
After computing the DPRT of the result along each direction, we can
      then take an inverse DPRT to recover $f$.

\begin{table*}
\newcommand{\meth}[2]{\multirow{#1}{0.10\textwidth}{#2}}
\newcommand{\blk}[2]{\multirow{#1}{  0.18\textwidth}{#2}}
\newcommand{\arch}[2]{\multirow{#1}{  0.23\textwidth}{#2}}
\newcommand{\alg}[2]{\multirow{#1}{  0.22\textwidth}{#2}}
\newcommand{\secc}[2]{\multirow{#1}{  0.15\textwidth}{#2}}
\caption{\label{table:overview}
Summary of the proposed methods.
Scalability is achieved by varying 
  the number of 1D convolvers ($J$) and 
  the number of rows processed in parallel in the DPRT and inverse DPRT ($H$).
There are some minor differences between the architectures that
  compute cross-correlations as opposed to convolutions.
Furthermore, convolution systems can implement cross-correlations
   by flipping the kernel offline or in real-time in hardware.
}
\begin{tabular}{llllll}
\toprule
   \meth{1}{\textbf{Method}}
 & \blk{1}{\textbf{Hardware Components}}
& \arch{1}{\textbf{Architecture}}
& \alg{1}{\textbf{Algorithm}}
& \secc{1}{\textbf{Section}}
\\ \midrule
  \meth{2}{\textit{FastConv}~/ \\
	          \textit{FastXCorr} }
 & \blk{2}{1D Circular convolver, \\
           FDPRT\cite{carranza2015}, iFDPRT\cite{carranza2015} \\
					 ~}
& \arch{2}{Fig. \ref{fig:1Dcirc_convo} (convolver), Fig. \ref{fig:SF2DLC} (system) \\
           using FDPRT and iFDPRT}
& \alg{2}{Fig. \ref{alg:conv1Dbasic} (convolver), Fig. \ref{alg:SF2DLC} (system)\\
           with $J=N+1$ and $H=N$}
& \secc{2}{\ref{sec:1DconvoCirc}, \ref{sec:F1DCC} and \ref{sec:ConvXross} \\
           ~}
\\ ~ \\ ~ \\
  \meth{2}{\textit{FastScaleConv}~/ \\
	          \textit{FastScaleXCorr} }
 & \blk{2}{1D Circular convolver, \\
           SFDPRT\cite{carranza2015}, iSFDPRT\cite{carranza2015} \\
					 ~}
& \arch{2}{Fig. \ref{fig:1Dcirc_convo} (convolver), Fig. \ref{fig:SF2DLC} (system) \\
           using SFDPRT and iSFDPRT}
& \alg{2}{Fig. \ref{alg:conv1Dbasic} (convolver), Fig. \ref{alg:SF2DLC} (system) \\
           Scale parameters: $J$ and $H$}
& \secc{2}{\ref{sec:1DconvoCirc}, \ref{sec:F1DCC} and \ref{sec:ConvXross} \\
           ~}
\\ ~ \\ ~ \\
  \meth{2}{\textit{FastRankConv}~/ \\
	          \textit{FastRankXCorr} }
 & \blk{2}{1D Linear convolver, \\
           Custom SRAM \\
					 ~}
& \arch{2}{Fig. \ref{fig:F1DLC} (convolver), Fig. \ref{fig:SRAMSystem} (SRAM),\\
           Fig. \ref{fig:S2DLCLUSystem} (system) }
& \alg{2}{Fig. \ref{alg:conv1Dsepbasic} (convolver), Fig. \ref{alg:S2DLCLU} (system) \\
           ~}
& \secc{2}{\ref{sec:LUarch} \\
           ~}
\\ ~ \\ 
\bottomrule
\end{tabular}
\end{table*}

\section{Methodology}\label{sec:methods}
We provide a summary of the proposed methods in Table \ref{table:overview}.
For general kernels, all methods are based on the DPRT and inverse DPRT.
Here, the fastest methods 
    (\textit{FastConv}, \textit{FastXCorr})
    correspond to a simplification of     
    the scalable methods
    (\textit{FastScaleConv}, \textit{FastScaleXCorr}).
For low-rank kernels, we recommend the use of \textit{FastRankConv} and 
    \textit{FastRankXCorr}.    
For the rest of the section,
    we begin with a description of the 1D convolver architecture
    that is shared by all methods (see section \ref{sec:1DconvoCirc}).
We then show how the 1D convolvers are integrated
    in the DPRT-based methods of sections
    \ref{sec:F1DCC} and \ref{sec:ConvXross},
    and the low-rank decomposition methods of section
    \ref{sec:LUarch}.
In section \ref{sec:overlap_add} we describe the application of overlap-and-add 
   for applying all of the methods 
    to large images.

\subsection{Computing 1D circular convolutions using circular shifts}\label{sec:1DconvoCirc}
Let $F_{m}(d), G_{m}(d), H_{m}(d)$ denote the DPRTs of $f, g, h$ along
       the $m$-th prime direction.
We define a special flip operation $\breve{H}_{m}$ defined by:
\[
   \breve{H}_{m}(d)= H_{m}(N-1-d), \quad d\geq 0,
\]
and the circular right shift (CRS) by $n$ using ${H}_{m}^n$
    that is defined by:
        \[
           {H}_{m}^n (d) = H_{m}(\left\langle d+n\right\rangle_{N}).
\]
Then, start from \eqref{eq:conv1dshort}
    to derive a shifted representation of the circular convolution using:
\begin{align}
    F_{m}(d)
         &= \sum^{N-1}_{k=0}G_{m}(k) \, H_{m}(\left\langle d-k\right\rangle_{N})  \nonumber \\
         &= \sum^{N-1}_{k=0}G_{m}(k) \, H_{m}(\left\langle N-1-k+d+1\right\rangle_{N}) \nonumber \\
         &= \sum^{N-1}_{k=0}G_{m}(k) \, H_{m}^{d+1} (N-1-k) \nonumber \\
         &= \sum^{N-1}_{k=0}G_{m}(k) \, {\breve{H}}_{m}^{d+1} (k). \label{eq:final1d}
\end{align}
From \eqref{eq:final1d}, we can see that  $F_{m}(d)$
   can be expressed as the dot product between $G_{m}$ and
   a flipped and circular right shifted by $d+1$ positions version of $H_{m}$
   (denoted as $\breve{H}_m^{d+1}$).

\subsection{Fast 1D circular convolution hardware implementation} \label{sec:F1DCC}
\begin{figure}[!bt]
\centering
\includegraphics[width=0.48\textwidth]{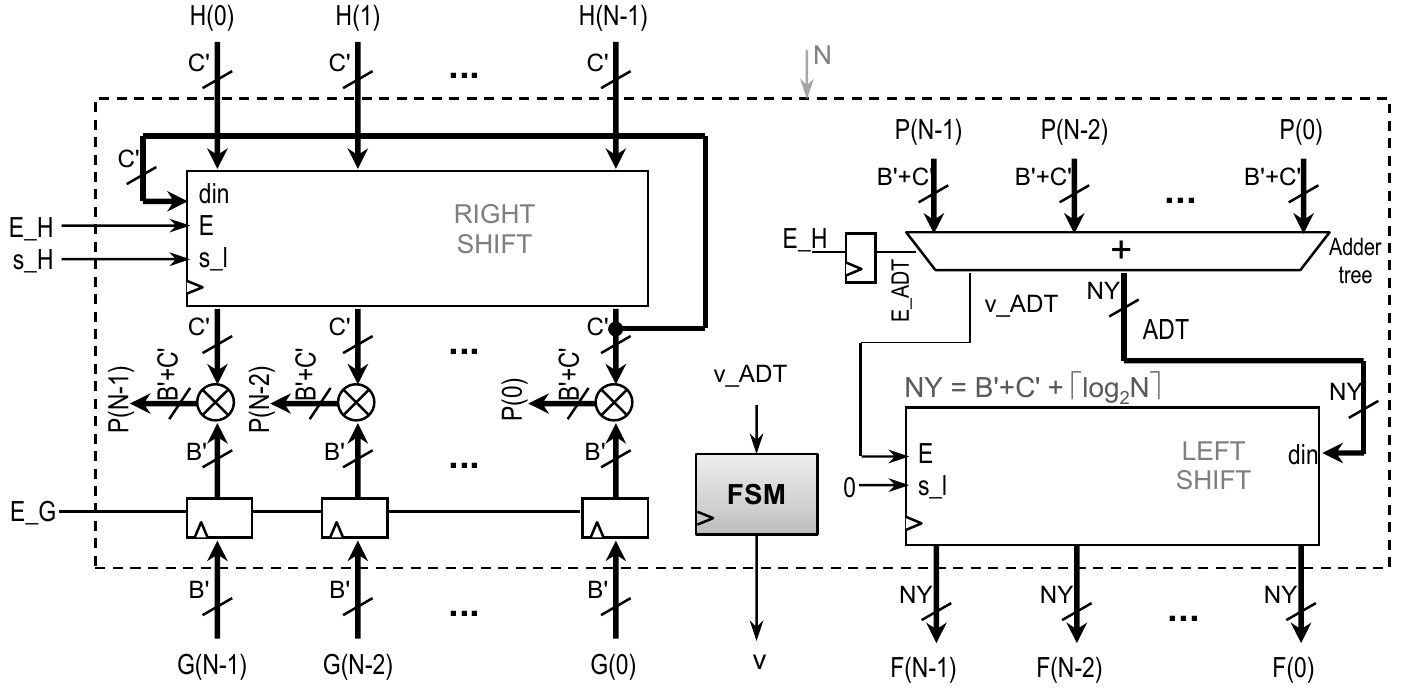}

\caption{
    Architecture for computing the 1D circular convolution $F_{m}= G_{m}\otimes H_{m}$.
    $B'$ and $C'$ represent the number of input bits  of $G$ and $H$ respectively.
}
\label{fig:1Dcirc_convo}
\end{figure}

\begin{figure}[!t]
\centering
\includegraphics[width=0.3\textwidth]{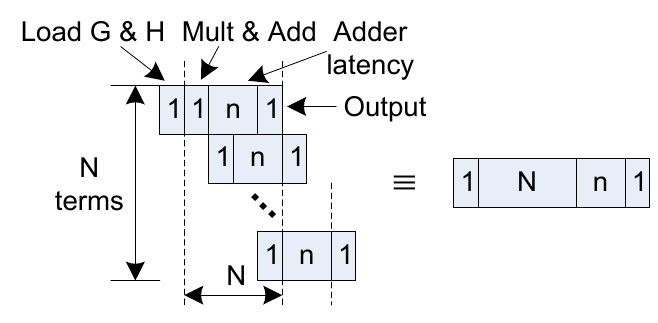}
\caption{Running time for the implementation of the fast architecture for computing
          one 1D circular convolution.
        In this diagram, time increases to the right.
        The number of clock cycles for computing each term of $F_{m}(d)$ is shown on each strip.
        The strip on the right represents the total running time.
        $n=\left\lceil \log_{2}N\right\rceil$ represents the addition latency.
}
\label{fig:1Dcirc_pipe}
\end{figure}

In this section, we derive a fast hardware implementation based on \eqref{eq:final1d}.
We present the hardware architecture in Fig. \ref{fig:1Dcirc_convo},
     the associated algorithm in Fig. \ref{alg:conv1Dbasic}, and the timing
     diagram in Fig. \ref{fig:1Dcirc_pipe}.

We begin with the fast computation of 1D circular convolutions given in Fig. \ref{alg:conv1Dbasic}.
Initially, we use parallel loads to transfer both of the DPRTs to the $G$ and $H$ registers
    in a single clock cycle.
Note that flipping $H_m$ into $\breve{H}_m$ is performed by simply wiring the inputs
    in reverse as shown in the upper register portion of Fig. \ref{fig:1Dcirc_convo}.
Starting with the last convolution output, we have a 3-step sequence of
    parallel multiplies, addition of the results, and a circular right shift
    to prepare for the next output
    (lines \ref{alg:conv1:mult}-\ref{alg:conv1:shift}).
The multiplications are performed in parallel in a single cycle using
     the parallel fixed-point multipliers of Fig. \ref{fig:1Dcirc_convo} and added
     using a pipelined tree structure in just $\lceil \log_2(N) \rceil$ clock cycles
     (e.g., see \cite{carranza2015}).
The resulting outputs are left-shifted in, one output sample at a time, into the
      output $F$ register shown in the lower-right portion of Fig. \ref{fig:1Dcirc_convo}.
A single cycle is also needed to perform the circular right shift of $H$
     using the top-left register of Fig. \ref{fig:1Dcirc_convo}.

To derive the timing requirements, refer back to Fig. \ref{fig:1Dcirc_pipe}.
Using a fully pipelined approach, we begin working on the next output sample
    after the parallel multiplies.
It is easy to see that after the initial latency for the first sample,
   we compute an output sample at every cycle.
After adding the latency for the initial loads, the adder latency,
   and the final left shift, we have a total
   of just $N+\lceil\log_2(N)\rceil+2$ clock cycles.

\begin{figure}[!tb]
\begin{algorithmic}[1]
\State Parallel load $G=G_m$, flipped load $H=\breve{H}_{m}$ \label{alg:conv1:load}
\For {$d$ = $N-1$ downto $0$} \label{alg:conv1:loop}
        \State Parallel mult.
              $P(k) = G(k)H(k), \,\, 0\leq k\leq N-1.$  \label{alg:conv1:mult}
              \vspace{0.02in}
        \State Parallel add
              $F(d) = \sum^{N-1}_{k=0}P(k)$  \label{alg:conv1:add}
              \vspace{0.02in}
        \State CRS by one $H={H}^1$  \label{alg:conv1:shift}
\EndFor  \label{step:conv1Dbasic7}
\State Parallel output $F$   \label{alg:conv1:out}
\end{algorithmic}
\caption{\label{alg:conv1Dbasic}
{ Algorithm for computing the 1D circular convolution $F_{m} = G_{m} \otimes H_{m}$.}
}
\end{figure}

\subsection{Fast and scalable 2D linear convolutions and cross-correlations
            using the DPRT}\label{sec:ConvXross}
In this section, we develop the architectures, algorithms, bit requirements,
   and computational efficiency for 2D convolutions and cross-correlations.
Most importantly, we discuss
    the scalability of the proposed approach that allows for the most efficient
    implementations based on available resources.

We begin with an analysis of the sequence of operations for computing fast
    and scalable 2D convolutions and cross-correlations as shown in Fig. \ref{alg:SF2DLC}.
In the most efficient implementation, the convolution kernel is available ahead of time.
In this case, we can pre-compute the DPRT of the kernel and store it in memory as shown in the
    hardware architecture of Fig. \ref{fig:SF2DLC}.
In Fig. \ref{fig:SF2DLC}, we provide a unifying architecture for implementing   
    \textit{FastScaleConv}, \textit{FastScaleXCorr},
    \textit{FastConv}, and  \textit{FastXCorr}.

For adaptive filterbank applications, the DPRT of the zero-padded
    convolution kernel can be computed in real-time 
    using the ${\tt SFDPRT\_System}$
    where the resulting DPRT is stored in (additional) memory. 
Alternatively, we can replicate the ${\tt SFDPRT\_System}$ system
            for the kernel to avoid an increase of the running time. 
For computing cross-correlations,
    we need to undo the vertical and horizontal
    flips associated with convolution.
This can be done by flipping the kernel
    along rows and columns
    as described in Fig. \ref{alg:SF2DLC}.
Here, note that the
    horizontal and vertical flips
    are performed by the ${\tt SFDPRT\_System}$ component
    during the loading of the kernel.
An inverted ${\tt MODE}$ signal 
    is used to control the ${\tt SFDPRT\_System}$
    to perform the needed flips.
Vertical flips are implemented by switching
    the order of loading the rows.
Thus, in a vertical flip,     
    the last kernel row is loaded first and
    the first kernel row is loaded last.
Horizontal flips are simply implemented
    by loading each row in reverse.
Thus, in a horizontal flip,
    the last row element is loaded first
    and the first row element is loaded last.
Overall, there is minimal overhead for implementing 
    the horizontal and vertical flips.

Scalability is achieved by controlling
  (i) the number of 1D circular convolutions that can be computed in parallel 
      (denoted by $J$), and
  (ii) the number of image rows that can be processed in parallel in the DPRT
        blocks (denoted by $H$ as described in \cite{carranza2015}).
Following the computation of the 1D circular convolutions, an inverse DPRT
   is applied for computing the final result.

We also list bit requirements.
For the setup, refer to the notation of section \ref{sec:notation}.
To compute exact convolutions, we need to zero pad to a prime number.
We thus require $N = {\tt NextPrime} (\max (P_1+Q_1-1, \, P_2+Q_2-1))$.
Then, it is easy to see that we require
  (i) $B+n$ bits for the DPRT of $g$, $C+n$ bits for the DPRT of $h$ where
      $g$ uses $B$ bits,
      $h$ uses $C$ bits, and
      $n=\left\lceil \log_2N\right\rceil$ (also see \cite{carranza2015}),
  (ii) $B+C+3n$ bits for the convolutions, and
  (iii) $B+C+4n$ bits just before the normalization step of the inverse DPRT 
        \cite{carranza2015},
   and $B+C+x$ bits for the final result, where $x$ represents the additional
            bits used for precision after the division.

\begin{figure}[!t]
\centering
\includegraphics[width=0.4\textwidth]{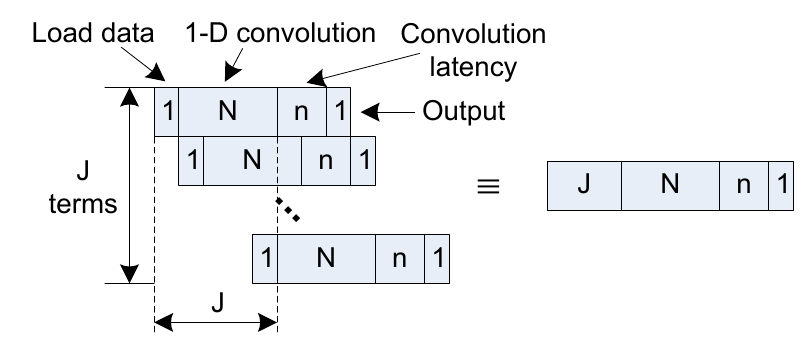}
\caption{Running time for computing $J$ circular convolutions in parallel
    using $J$ fast convolution blocks (see basic block structure in Fig. \ref{fig:1Dcirc_convo}).
In this diagram, time increases to the right.
Here, it takes one cycle to perform a parallel load for each block.
Overall, we require $J+N+n+1$ to compute everything, where
        $n=\left\lceil \log_{2}N\right\rceil$ represents the addition latency.
}
\label{fig:J1Dcirc_pipe}
\end{figure}

\begin{figure}[!t]
\centering
\includegraphics[width=0.45\textwidth]{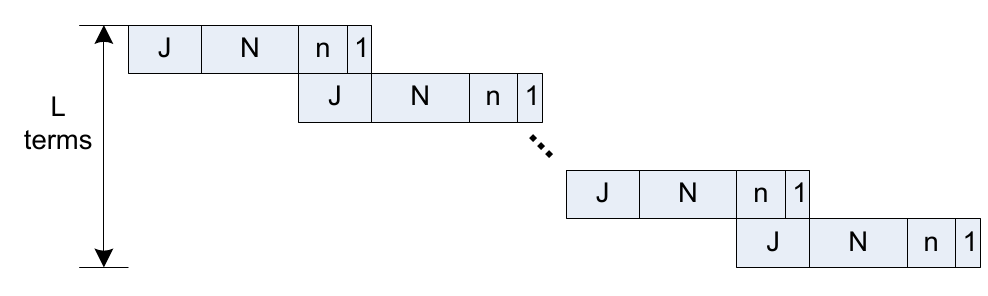}
\caption{
Running time for computing $N+1$ 1D circular convolutions using
       $J$ fast convolution blocks operating in parallel.
In this diagram, time increases to the right.
We need to reload the convolution blocks $L$ times given
      by $L=\left\lceil (N+1)/J \right\rceil$.
Each row shows the running time for performing $J$ convolutions
     as described in Fig. \ref{fig:J1Dcirc_pipe}.
}
\label{fig:N1Dcirc_pipe}
\end{figure}

We next derive the computational complexity of our approach.
From section \ref{sec:dprt-only} and \cite{carranza2015},
     scalable DPRT computation requires
     $\left\lceil N/H\right\rceil(N+3H+3)+N+
      \left\lceil \log_{2}H\right\rceil+1$ clock cycles
     that reduce to
     $2N+\left\lceil \log_2N\right\rceil+1$ clock cycles
     for the fast DPRT implementation.
For computing the number of cycles required for the circular
   convolutions, refer to Figs.  \ref{fig:J1Dcirc_pipe}  and \ref{fig:N1Dcirc_pipe}.
As shown in Fig. \ref{fig:J1Dcirc_pipe},
    we require $J+N+n+1$ clock cycles to compute
    $J$ convolutions in parallel where
        $n=\left\lceil \log_{2}N\right\rceil$ represents 
        the initial addition latency.
To compute outputs for all
    of the $N+1$ required DPRT directions,
   we use all $J$ parallel blocks of 1D convolutions for
   $L=\lceil (N+1)/J \rceil$ times. 
Depending on $N$, increasing $J$ may not always provide for
   better solutions.
There is a need to find optimal values for $J$.
We refer to section \ref{subsec:pareto} for determining
   the optimal values of $J$.
Overall, we require a total of 
     $L\cdot(J+N)+n+1$ clock cycles
     to compute all of the 1D convolutions.
We provide a summary of the required resources
    for implementing the $J$ 1D parallel convolution blocks in Table \ref{table:resources}.

Overall, based on the derived complexity, we have
     the fastest running time 
     using $J=N+1$ parallel 1D convolutions
     at just $2N+n+2$ clock cycles with 
     resource usage of $O(N^{2})$
     for flip-flops and full adders.
For the minimum number of resources,
     we only use a $J=1$ 1D convolution block
     that require
     $(N+1)^{2}+n+1$ clock cycles 
     with the lowest resource usage $O(N)$
     for flip-flops and full adders.

Following the 1D convolutions,
     we take the inverse DPRT using the ${\tt iSFDPRT\_System}$ module.
Similar to the forward DPRT, scalability is controlled by $H$,
       the number of image rows processed in parallel \cite{carranza2015}.
For this step, 
    the input data uses $B+C+3n$ bits per pixel.
Depending on available resources, the inverse DPRT can
     be computed in just $2N+5n+B+C+2$ for the fast inverse DPRT 
     with $O(N^{2})$ resource usage (1-bit additions and flip-flops), 
     or as slow as
     $\left\lceil N/2\right\rceil(N+2)+4n+B+C+4$  for $H=2$
     for just $O(N)$ resource usage \cite{carranza2015}.

\begin{figure}[!tb]
\algnewcommand{\LineComment}[1]{\Statex \(\triangleright\) #1}
\begin{algorithmic}[1]
\LineComment{For cross-correlation, apply:}
\LineComment{$\quad$ {\tt Flip} ($h$) and store }
\LineComment{$\quad$ the flipped version in memory.}
\vspace{0.03 true in}
\State Precompute/Compute   \label{alg:conv:first}
\Statex $\quad H = {\tt DPRT}\{{\tt ZeroPad}\{h\}\}$
\Statex $\quad$ and store the results in memory.
\vspace{0.03 true in}
\State Compute $G = {\tt DPRT}\{{\tt ZeroPad}\{g\}\}$
\vspace{0.03 true in}
\For {$p$ = $0$ to $L-1$}
        \State Compute $J$ directions in parallel:
        \Statex $\qquad F_{pJ+i} = G_{pJ+i} \otimes H_{pJ+i},
                       \quad \text{for} \,\, i=0, \ldots , J-1.$
\EndFor  \label{step:SF2DLCstep1}
\State Compute $f={\tt DPRT}^{-1}\{F\}$
\end{algorithmic}
\caption{\label{alg:SF2DLC}
 Fast and scalable algorithm for computing 2D linear convolutions
      and cross-correlations
      between $g(i,j)$ and $h(i,j)$ using the
     architecture depicted in Fig. \ref{fig:SF2DLC}.
		$L=\lceil (N+1)/J \rceil$
  }
\end{figure}

\begin{figure*}[!bt]
\centering
\includegraphics[width=1.0\textwidth]{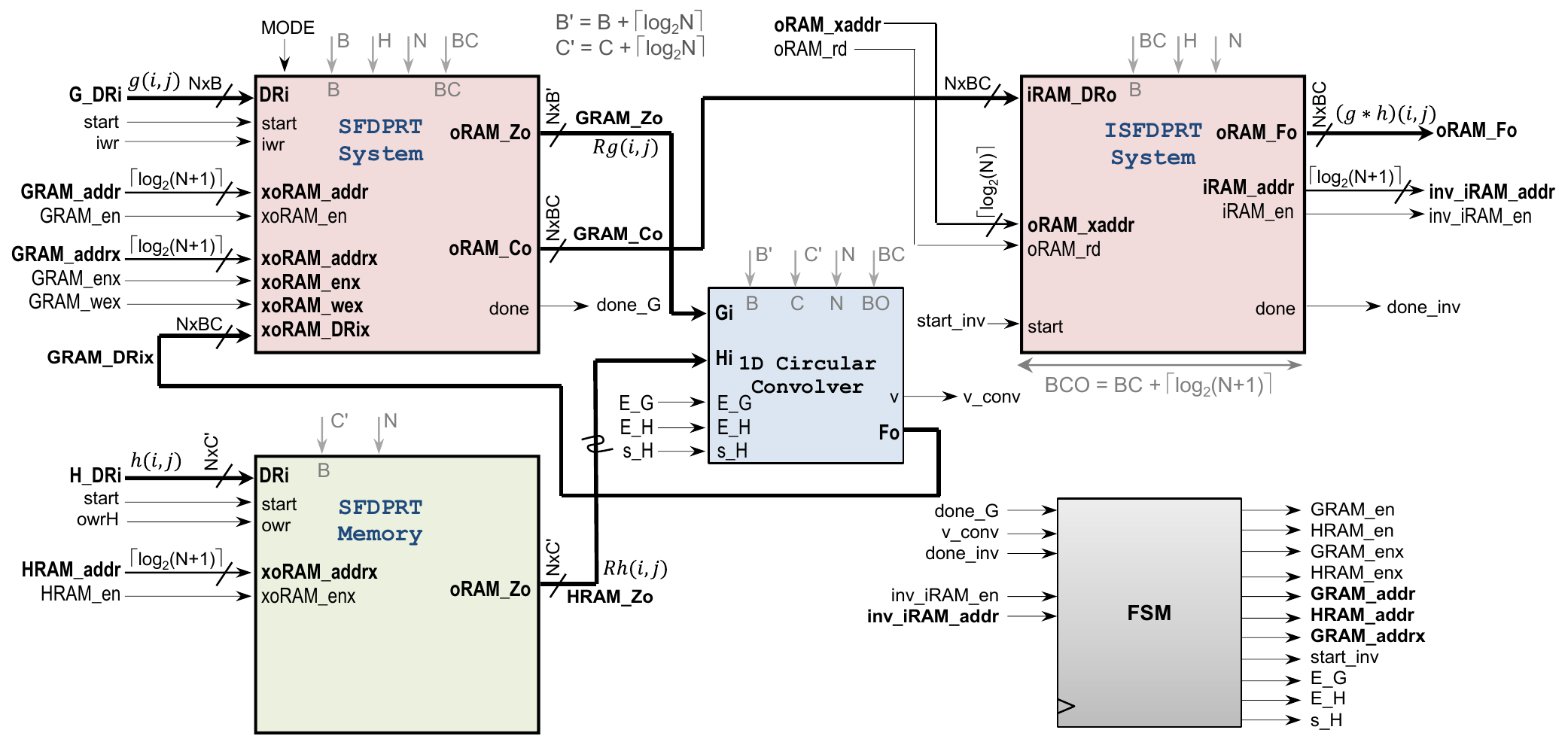}
\caption{
\textit{FastScaleConv} and \textit{FastScaleXCorr}:
   Fast and scalable architecture system for 
      computing 2D convolutions and cross-correlations 
      based on the DPRT (also see Fig. \ref{alg:SF2DLC}).
The forward DPRT is computed by SFDPRT.
The inverse DPRT is computed by ISFDPRT.
The linear convolver computes circular convolutions for $J$ rows.
The finite state machine (FSM)  
    manages all the control signals 
    (except for 'start' and 'iwr'). 
We use bold face letters to denote buses
    while convolution parameters are depicted in gray.
Refer to section \ref{sec:notation} for definitions of the basic convolution parameters.
\textit{FastConv} is a simplification of \textit{FastScaleConv} for maximum performance (see text).
   }
\label{fig:SF2DLC}
\end{figure*}

\subsection{Fast and scalable 2D linear convolution using SVD-LU decompositions
                   (FastRankConv)} \label{sec:LUarch}
As described in section \ref{sec:LUdecomp}, we can use
   a collection of 1D convolutions along the rows and columns 
   to implement effective approximations to 2D convolutions
	with the inherent loss in accuracy due to zeroing the smaller singular values 
	associated with the SVD decomposition.
Unfortunately, direct approaches suffer from the need to
   implement two transpositions that require $O(N^2)$ clock cycles. 
In this subsection, we present a fast and scalable system that
   eliminates the need for transpositions and allows us to
   compute convolutions in $O(N)$ to $O(N^2)$ clock cycles with the addition of an intermediate  custom memory.

\begin{figure}[!t]
\centering
\includegraphics[width=0.5\textwidth]{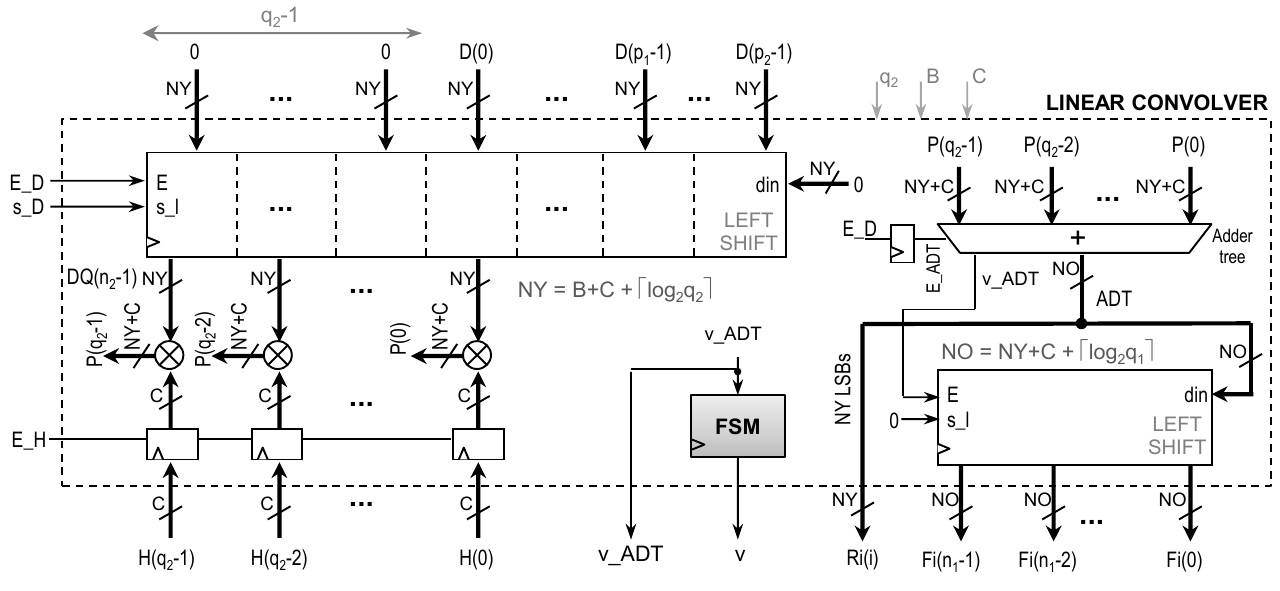}
\caption{
 Architecture for computing fast 1D linear convolution $F_{i}= D_{i}\ast H_{i}$.
    This block is the core for computing 2D convolutions with separable kernels.
		It computes 1D convolutions for two sizes:
		  (i) $D$ and $H$ are of size $p_2$ and $q_2$ respectively,
		      generating outputs of size $p_2+q_2-1$, performed $p_1$ times (the rows of the 2D image or image block).
		 (ii) $D$ and $H$ are of size $p_1$ and $q_1$ generating outputs of size $p_1+q_1-1$,
		      performed $p_2+q_2-1$ times (the columns of the step (i) result).
		Refer to Fig. \ref{fig:S2DLCLUSystem} (bottom part) for more details about the 2D convolution.
}
\label{fig:F1DLC}
\end{figure}

As before, scalability is achieved by controlling $J$,
   the number of linear convolutions that are computed in parallel.
The linear convolution blocks are similar to the circular convolution
   blocks except that the complexity is a function of the size
   of the convolution kernel only (see Fig. \ref{fig:F1DLC} and bottom part of Fig. \ref{fig:S2DLCLUSystem}).   
Then, in order to operate as fast as possible, we design
   a custom memory system that moves entire rows or columns
   to and from each linear convolver.
The basic idea is to start by moving all of the rows into the 
    $J$ convolution blocks,
    store the convolution results in $J$ SRAM memories so
    that the rows of the row-convolutions results correspond
    to the columns of the original image, 
    and then perform row convolutions and store
    in $J$ output SRAM memories.
Thus, the need for the transpositions is completely avoided.     

\begin{figure}[!t]
\centering
\includegraphics[width=0.5\textwidth]{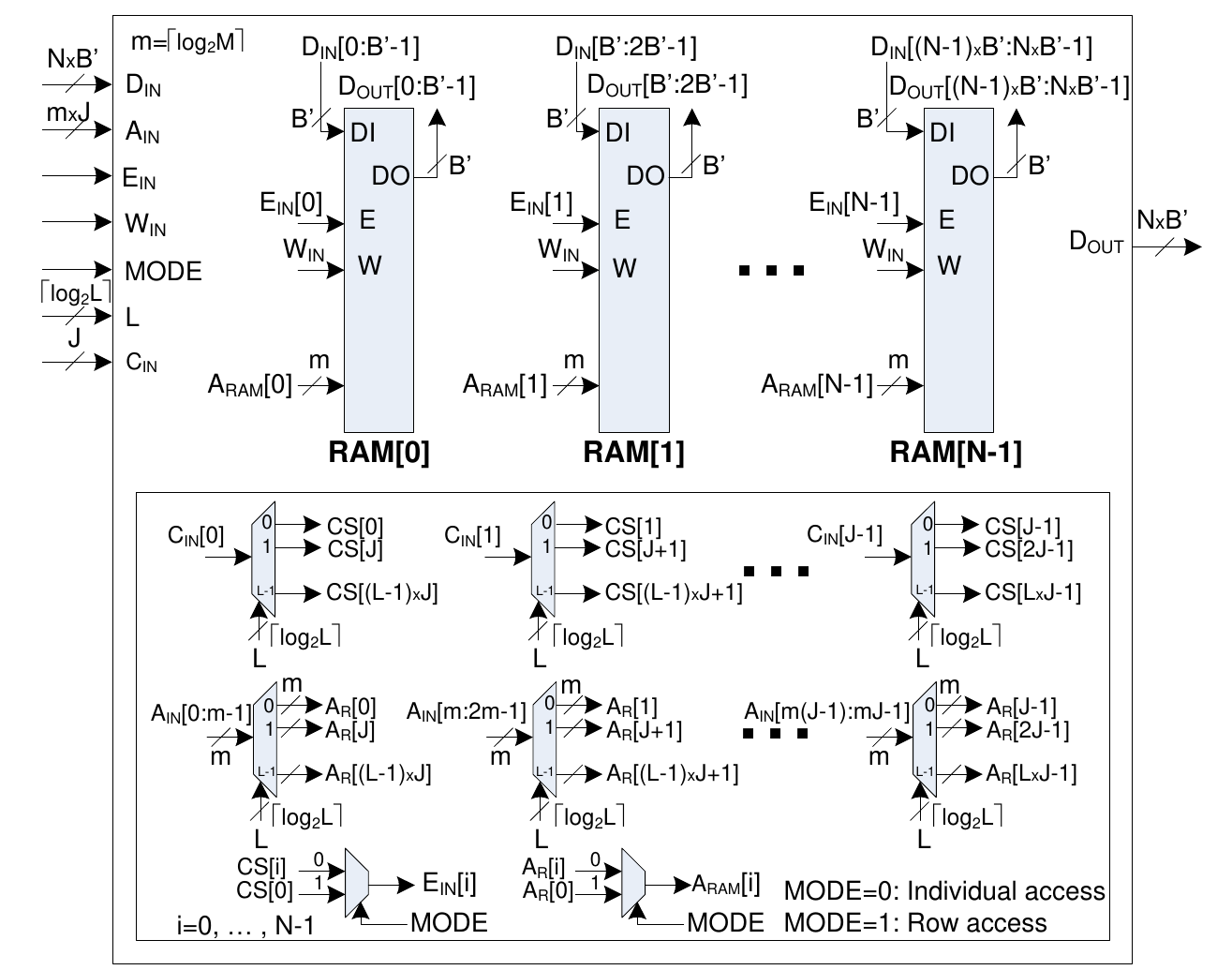}
\caption{
Custom SRAM architecture for fast transpositions and memory access.
The architecture allows 
   for full-row (or full-column, i.e. transpose) 
   read/write in a single clock cycle ({\tt MODE=1})
   and individual access to up to $J$ SRAMs in a single clock cycle ({\tt MODE=0}).  
The SRAM stores $M$ rows (or columns) of $N$ $B'$-bit per pixels.
The basic architecture can be configured for different purposes as given in
   Table \ref{table:MEM_config}.
}
\label{fig:SRAMSystem}
\end{figure}

\begin{figure*}[!t]
\centering
\includegraphics[width=1.0\textwidth]{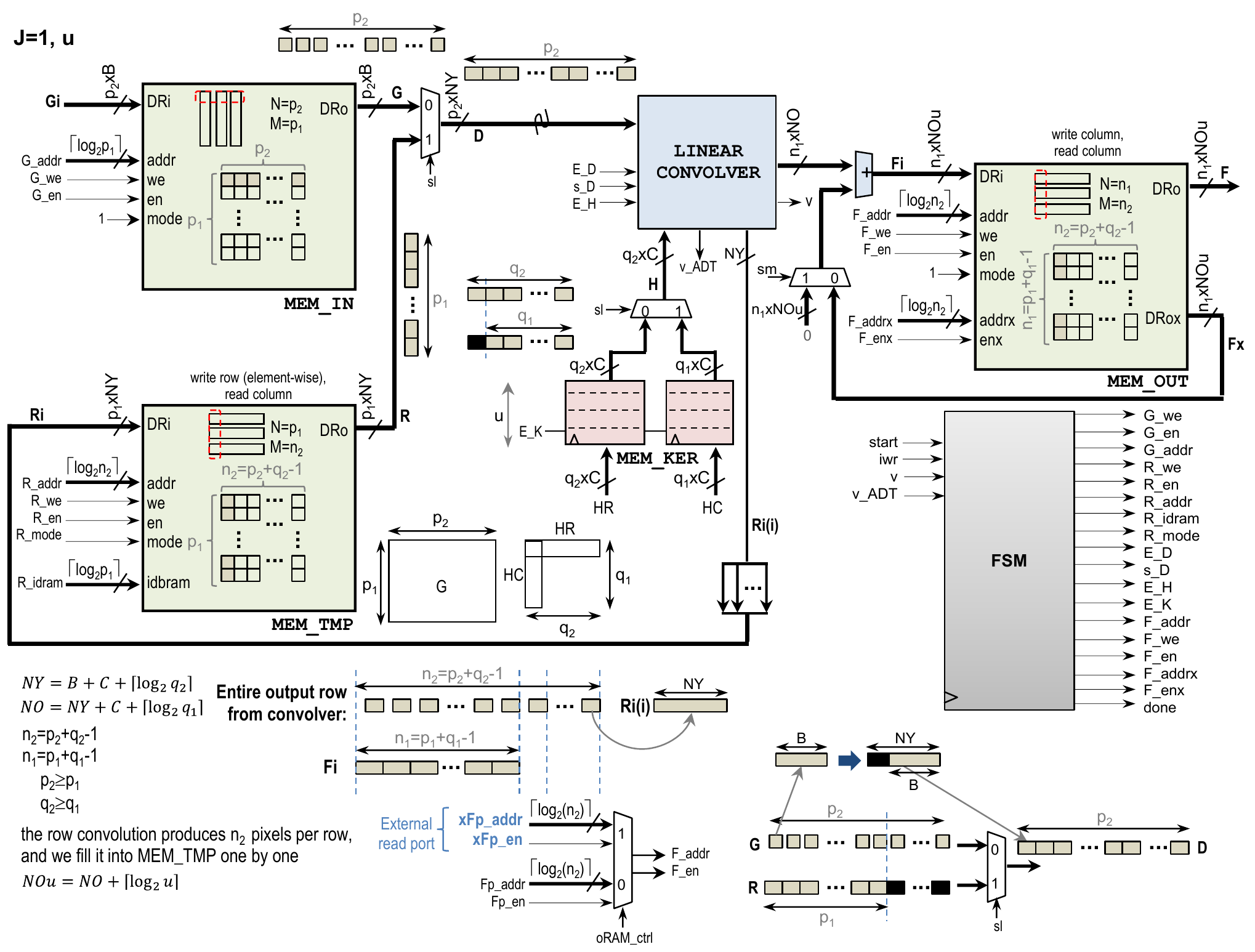}
\caption{
 \textit{FastRankConv}: Fast and scalable 2D convolution system based on separable decompositions.
 Refer to subsection \ref{sec:notation} for the notation
      and Fig. \ref{fig:SRAMSystem} for the memory architecture.
 The linear convolution blocks are very similar to the
      circular convolution blocks 
      except that multiplications and additions are reduced to
      the size of the convolution kernel.
 The Bus width shown is the one for maximum accuracy. 
 Also, note that the implementation of \textit{FastRankCross} 
     is not considered here since cross-correlation is the same as convolution with a flipped kernel, 
     and flipping can be computed during pre-processing (prior to SVD and LU). 
The finite state machine is denoted by FSM.
The Linear Convolver can be applied to several rows in parallel.
The results are accumulated in $\texttt{MEM\_OUT}$.
Please refer to Fig. \ref{alg:S2DLCLU} for a full description of the 
    algorithm and section \ref{sec:notation} for definitions
    of the basic parameters (p1, q1, p2, q2, B, C).
}
\label{fig:S2DLCLUSystem}
\end{figure*}

Then, for a single clock cycle,
   we use custom memories to 
   (i) allow us to move entire rows and columns of blocks of pixels 
       from memory to the convolution blocks and vice-versa, and
   (ii) allow direct access to $J$ different SRAMs.
We present the proposed custom SRAM architecture in Fig. \ref{fig:SRAMSystem},
   the full system architecture in Fig. \ref{fig:S2DLCLUSystem} and
	 the associated algorithms in Figs. \ref{alg:conv1Dsepbasic} and \ref{alg:S2DLCLU}.
Refer to section \ref{sec:notation} for the notation.  
We customize the basic SRAM architecture of Fig. \ref{fig:SRAMSystem}
  as given in Table \ref{table:MEM_config}, so that in a single clock cycle: 
  (a) ${\tt MEM\_IN}$  provides a full row of the image,
  (b) ${\tt MEM\_KER}$ provides the entire row or column filter
      coefficients,
  (c) ${\tt MEM\_TMP}$ stores the results of convolution along each row,
      provides access to a full column of the results, and
  (d) ${\tt MEM\_OUT}$, accumulates the final result, 
      adds up to $P2+Q2-1$ values of the convolved image (in a single clock cycle), and 
      also provides a full row.      
The required resources are summarized in Table \ref{table:resourcesLU}.

We also provide a summary of performance-resource requirements.
Without loss of generality, we assume that $P2 \geq P1$, $Q2 \geq Q1$, 
   and consequently $N2 \geq N1$.
Furthermore, for the purposes of the analysis,
   assume full rank: $r=Q1$,
   and let 
   $L_R = \left\lceil P1/J\right\rceil$ and
   $L_C = \left\lceil (P2+Q2-1)/J\right\rceil$.
The total running time is given as the sum
   of clock cycles required for:
   (i)   row processing:    $r\cdot L_R\cdot (J+P2+Q2-1)$,
   (ii)  column processing: $r\cdot L_C\cdot (J+P1+Q1-1)$, and
   (iii) the latency of the adder tree $\left\lceil \log_2 Q1\right\rceil + 1$.
To simplify the derivation, let   
    $N=\max\left\{P2+Q2-1,P1+Q1-1\right\}$.
Then,
 for $J=1$, we have minimum resource usage that grows as $O(N)$ 
    with a running time of $O(N^2)$.
For $J=N$, we have the fastest running time $O(N)$ with resource usage that grows as
    $O(N^2)$.
Refer to Table \ref{table:MEM_config} for detailed resource usage of the memories.
We can further optimize the architecture parameters as described in
 section \ref{subsec:pareto}.
We will also provide more detailed comparisons in 
 section \ref{sec:results}.

\begin{table}
\caption{\label{table:MEM_config} 
  SRAM memory configurations for maximum accuracy.
  Orientation refers to each SRAM holding either a full row or column of the image.
  The Accumulate mode needs external adders to perform the accumulation and
     dual-port SRAMs for full speed.
  $B$ denotes the number of bits of the input image.
  $C$ denotes the number of bits used for the kernel coefficients.
  We have
     $q1=\left\lceil \log_2 Q1\right\rceil$ and
     $q2=\left\lceil \log_2 Q2\right\rceil$.
}
\begin{center}
\resizebox{0.5\textwidth}{!}{%
\begin{tabular}{p{0.07\textwidth}llll}
 \toprule 
 SRAM & ${\tt MEM\_IN}$ & ${\tt MEM\_KER}$ & ${\tt MEM\_TMP}$ & ${\tt MEM\_OUT}$ \\
 \midrule 
 Quantity       &       $P2$ & $Q2$ & $P1$ &    $P2+Q2-1$  \\
 Depth & $P1$ & $2Q2$ & $P2+Q2-1$ & $P1+Q1-1$  \\
 Wordlength & $B$ & $C$ & $B+C+q2$ & $B+2C+q1+q2+r$  \\
 Function &     $g$ & $h_R,h_C$ & $g'$ &        $f$  \\
 MODES &        $1$ & $1$ & $0/1$ &     $0/1$  \\ 
 Orientation & Column & Column & Row & Column \\
 WriteMode & Store & Store & Store & Accumulate \\
 \bottomrule 
\end{tabular}}
\end{center}
\end{table}

\begin{figure}[h]
\begin{algorithmic}[1]
\Procedure {{\tt LinConv1D}}{$D,SG,SH$, \texttt{MEM}}
\State Parallel load $GX[SH-1:SG-1] = D$ \label{step:conv1Dsepbasic1}
\State Parallel load $GX[0:SH-2] = 0$
\For {$s$ = $0$ to $SG-1$} \label{step:conv1Dsepbasic3}
        \State Parallel mult. $P(k) = GX[k]\, H[k]$  \label{step:conv1Dsepbasic4}
        
        \ \ \ \ \ \ \  for $k=0, \ldots , SH-1$
        \State Parallel add $F[s] = \sum^{SH-1}_{j=0}P[j]$ \label{step:conv1Dsepbasic5}
        
        \ \ \ \ \ \ \  and store or accumulate in \texttt{MEM}
        \State CLS by one $GX$  \label{step:conv1Dsepbasic6}
\EndFor  \label{step:conv1Dsepbasic7}
\EndProcedure
\end{algorithmic}
\caption{\label{alg:conv1Dsepbasic}
  Algorithm for computing 1D linear convolution between
      $D$ and the 1D kernel $H$   (size=$SH$)
      and stores the results in $F$ (size=$SG$).    
  CLS refers to the circular left shift operation.
  $GX$ represents the upper-left row of shift registers in Fig. \ref{fig:F1DLC}.
  $HX$ represents the lower-left row of registers in Fig. \ref{fig:F1DLC}
     that is pre-loaded with the 1D kernel ($H$).
  The output is stored in ${\tt MEM = MEM\_TMP}$ for rows, or accumulated in ${\tt MEM = MEM\_OUT}$ 
    for columns. 
}
\end{figure}

\begin{figure}[h]
\begin{algorithmic}[1]
\For {$q$ = $0$ to $r-1$}
\Statex $\quad$ \(\triangleright\) Compute convolutions along the rows
\State  Parallel load row-kernel using
\Statex \ \ \ \ \ \ \ \ \ \ \ \ \ \ $HX[0:Q2-1]= h_{Rq}(j)$
\For {$p$ = $0$ to $L_R-1$}
\State Use $J$ 1D-convolvers to compute: 
\Statex \ \ \ \ \ \ \ \ \ \ \ \ {\tt LinConv1D} ($g_{p\cdot J+k}(j)$, 
        
        \ \ \ \ \ \ $\qquad\qquad\qquad$ $P2+Q2-1$, $Q2$,
        
        \ \ \ \ \ \ $\qquad\qquad\qquad$ ${\tt MEM\_TMP}$)
        
        \ \ \ \ \ \ \  for $k=0, \ldots , J-1$.
\EndFor
\Statex $\quad$ \(\triangleright\) Compute convolutions along the columns
\State Parallel load column-kernel using
\Statex \ \ \ \ \ \ \ \ \ \ \ \ \ \ $HX[0:Q1-1]= h_{Cq}(j)$
\For {$p$ = $0$ to $L_C-1$}
\State Use $J$ 1D-convolvers to compute: 
\Statex \ \ \ \ \ \ \ \ \ \ \ \ {\tt LinConv1D} ($g'_{p\cdot J+k}(i)$, 

        \ \ \ \ \ \ $\qquad\qquad\qquad$ $P1+Q1-1$, $Q1$, 
        
        \ \ \ \ \ \ $\qquad\qquad\qquad$ ${\tt MEM\_OUT}$)
        
        \ \ \ \ \ \ \ for $k=0, \ldots , J-1$
\EndFor
\EndFor
\end{algorithmic}
\caption{\label{alg:S2DLCLU}
Algorithm for computing the 2D linear convolution between an image (or image block) $g(i,j)$
    and the non-separable kernel $h(i,j)$ decomposed into 
    $r$ separable kernels.
We use $h_R(i,j)$ for the row kernels and $h_C(i,j)$ for the column kernels.
The results are computed in $g'(i,j)$.
The row-convolution results are stored in ${\tt MEM\_TMP}$.
The final output is accumulated in ${\tt MEM\_OUT}$.
}
\end{figure}

\subsection {Scalability for Large Images Using Overlap-Add}\label{sec:overlap_add}
The convolution and cross-correlation kernels tend to be much smaller than
       the size of the input image.
Thus, for much larger images, the best approach is to design
      the hardware architecture for the smaller kernels.
We summarize the basic approach below.

The original image is subdivided into the smaller windows
      that are equal to the size of the kernel.
Convolutions and cross-correlations can be computed
      for each block.
Results from neighboring blocks must be added together.
Furthermore, the final output is a concatenation of
      the results from each block.

The basic approach is very well understood.
Furthermore, the approach can also be parallelized 
      to use multiple hardware blocks.
In what follows, we will simply assume that both
     the image (block) and the convolution/cross-correlation kernel size
     are of the same size.
Furthermore, we will focus on the most common size when both
    the image (block) and the kernels are square.               

\subsection{Pareto optimal architectures}\label{subsec:pareto}
As we discussed earlier in this section, it is possible
   to use $J$ that is sub-optimal.
Here, we refer to an architecture as being sub-optimal
   in the Pareto sense 
   (e.g., see \cite{Llamocca2015}).
Essentially, an architecture is considered to be Pareto-optimal
   if it provides the best possible performance for required resources.
Thus, a Pareto optimal family of architectures      
   will always produce better running time
   for more resources.
To derive the set of Pareto-optimal solutions,
   recall that our scalable families of architectures
   may contain less than $J$ rows for the last block of 1D convolutions.
Thus, for \textit{FastScaleConv} and \textit{FastScaleXcross},
   to fully utilize available hardware resources,
   we require that the selected $J$ values would
   satisfy
   $\left\langle N+1 \right\rangle_{J}=0$.
Similarly, for \textit{FastRankConv},
   we require that the selected $J$ values simultaneously
   satisfy $\left\langle P1\right\rangle_J=0$ and
   $\left\langle P2+Q2-1\right\rangle_J=0$.

\section{Results}\label{sec:results}
In this section, we provide extensive 
   comparisons with prior methods to 
   demonstrate the promise of the proposed methods.
Here, we note that the proposed systems
   implement both convolutions and cross-correlations.

We compare our approach to 
   relevant, state of the art, convolution
   systems by considering
    (i) serial systolic arrays \cite{Hon1990} (SerSys),
   (ii) scalable and parallel systolic arrays \cite{Mohanty1996} (ScaSys),
  (iii) sliding windows \cite{Cooke2015} (SliWin), and
   (iv) parallel and pipelined Fast Fourier Transform radix-2 
        \cite{Ayinala2012} (FFTr2).
Here, we do not consider methods based on 
    Distributed arithmetic (DA) solutions 
    since the internal ROM required for the DA operation
    grows exponentially with the kernel size, making them 
    unsuitable for large kernels \cite{Meher2008}.
Furthermore, to provide fair comparisons,
    we are assuming that FFTr2 is based on 
    the parallel use of the highly efficient 1D
    FFTs described in \cite{Ayinala2012}.

We will consider comparisons for different bitwidths.
Here, we note that the required number of bits for maintaining
   full precision was developed in section \ref{sec:methods}.
Furthermore, we will provide results from full precision,
   the use of DSPs, and a limited bitwidth in
   the results.   
We are currently researching different methods for
   selecting different bitwidths for different
   stages of the algorithms.

As described earlier, the proposed architectures
   can compute both convolutions and cross-correlations.
For \textit{FastRankConv}, flipping the kernel
   can clearly 
   be done during pre-processing, prior to SVD and 
   LU computations.
In what follows, we will present
   results for 
   \textit{FastConv}, \textit{FastScaleConv},
   and \textit{FastRankConv}.
Here, we note that 
   \textit{FastXCorr}, \textit{FastScaleXCorr},
   and \textit{FastRankXCorr} 
   are minor variations of
   \textit{FastConv}, \textit{FastScaleConv},
   and \textit{FastRankConv}.
   
We describe the implementation setup in 
   section \ref{sec:setup}.
In section \ref{sec:setup}, 
   we also describe 
   alternative methods.
We provide extensive comparisons in terms
   of performance and required hardware resources
   in section \ref{sec:multiObjComps}.
FPGA and SOC implementations are described
   in section \ref{sec:fpga}.
   
\subsection{Implementation setup}\label{sec:setup}   
We consider convolutions with $P\times P$ kernels and image blocks where the output
   is of size $N\times N$ where $N=2P-1$.
For section \ref{sec:multiObjComps},
   we assume 
   $B=8$  bits for the input image pixels and 
   $C=12$ bits for the kernel coefficients.
We use $12$-bits for the outputs of the 
   additions, multiplications, and the DPRT
   of section \ref{sec:multiObjComps}.
We consider $C=8$ bits for the kernel coefficients
   and full-precision for the outputs in the FPGA and SOC implementations
   of section \ref{sec:fpga}.        
For the FFTr2, the computations are performed using
   32-bit floating point units.
  
For alternative image representations,
   we briefly refer to 
   Table \ref{table:ResultsResources}
   and   Fig. \ref{alg:adderResources}
   that are covered in more detail in   
   section \ref{sec:multiObjComps}.
In Fig. \ref{alg:adderResources} we provide
   an algorithm for computing the 
   (i) number of flip-flops inside the adder tree and
   (ii) the equivalent number of 1-bit additions
   as functions of the number of bits per pixel $D$
   and the size of the final output image $N$. 
From Fig. \ref{alg:adderResources}, we can see
   that increasing the number of input bits from 8 to 24 will
   linearly increase the numbers of adder tree flip-flops and 1-bit additions
   but remain bounded above by $3\times$ the amounts
   presented here.
In terms of the fixed-point multipliers, 
   at 24-bits, we would also consider
   the use of highly-optimized, 32-bit floating point units.  
Furthermore, we note that there will be a minimal impact on running time.
Overall, we note that our fixed-point implementations are
   most effective for the most commonly used, 8-bit inputs.
  
To enable comparisons between FFTr2 and fixed-point implementations,
   we make some simple, and realistic approximations for 
   possible FPGA and SOC implementations.
Based on Tables 2 and 3 from \cite{Vera2011},
     in terms of 1-bit adders, we have that 
     32-bit floating point additions can be approximated as 10$\times$ 
     the cost of 32 1-bit fixed-point additions.
Furthermore, to compare FFTr2 to all other implementations,
     we need to compare
     resources for 32-bit floating-point multiplication
     in terms of resources for 12-bit fixed-point multiplication.
To provide realistic comparisons,
    we implemented complex, 32-bit floating point multiplications on
    the Virtex-7 (XC7VX1140).
Without using DSPs, the multipliers require
    650 LUTs and 32 flip-flops.
When using DSPs, the multipliers require
    34 LUTs, 32 flip-flops, and 2 DSPs.
On the same FPGA, without using the DSPs, 
    a 12-bit fixed point multiplier required just 148 LUTs.
In terms of LUTs, we have: $650/148 \approx 4.4$.
We thus approximate the cost
  of 32-bit floating point multiplications as
  equivalent to $4.4$ times the cost of 12-bit fixed point multiplications.

For FFTr2, \cite{Ayinala2012} does not provide detailed running time
   for the complete convolution of 2D images.
To provide for fair comparisons, we consider an extension
   of FFTr2 using point-to-point multiplications
   using $D$ 1D FFT cores.
Then, in the fastest possible 2D implementation,
   we assume that it would take $N^2/D$ additional clock cycles
   to implement the point to point complex multiplications.
       
As discussed earlier, for \textit{FastScaleConv},
   we achieve hardware scalability by
   varying $H$, the number of rows processed in parallel for the scalable DPRT,
   and $J$, which represents the number of 1D convolutions computed in parallel.
Here, for $H=2, 3, \dots, N-1$, 
   we will simply set $J=H$ for a balanced approach towards both.
Then, for $H=N$, we use $J=N+1$ 
   to provide the optimal solution using \textit{FastConv}.
For \textit{FastRankConv}, we use $r$ to denote the rank of the approximation.
   
There are some special restrictions on $N$.
For the DPRT-based methods,
    $N$ needs to be prime.
For FFTr2, $N$ is assumed to be a power of 2.
For ScaSys, $P$ needs to be a composite number ($N=2P-1$),
    and we thus assume that $P=P_A \cdot P_B$.
We focus on the cases when 
    $P_A=2$ (slowest) and $P_B=4$ (fastest),
    with an input buffer and fully pipelined additions.
We do not include the case when
    $P_B=2$ because the resource usage becomes prohibitive ($O(N^3)$).
For SerSys, SliWin and 
    \textit{FastRankConv}, 
    there is no restriction for $P$.
When needed to change the size, we apply zero-padding.

\subsection{MultiObjective Comparisons}\label{sec:multiObjComps}   
A primary contribution of the manuscript is to provide
  convolution and cross-correlation architectures that are both fast and scalable.
Because of scalability, for most reasonably-sized devices and convolution sizes,
  we can find possible implementations that can
  fit within it.
On the other hand, there is great variability 
  among different devices.
Here, for an expanding range of $P$ values,
  we show that our proposed architectures
  are optimal in the multi-objective sense.
In other words, for any given level of fast performance
   (from $O(P)$ up to $O(P^2)$ clock cycles),
   the required architectures require fewer hardware resources.        

In terms of transform size scalability,
  we note that there are several prime numbers
  between any two of powers of two.     
For example, 
   from $4$ to $256$, we only have 7 possible
   sizes for FFTr2, compared
   to $53$ prime numbers
   for \textit{FastConv} and \textit{FastScaleConv},  
   and no size restrictions for \textit{FastRankConv}.
As a result, for $P=65$, $N=129$ requires
   zero-padding to $256$ for FFTr2 while
   it can be handled by a $131$-sized
   DPRT associated with
   \textit{FastConv} and \textit{FastScaleConv}.
   
We present a comprehensive summary in terms of performance
   and resources in Table \ref{table:ResultsResources}.
Table \ref{table:ResultsResources} 
   lists performance in clock cycles,
   number of flip-flops, 
   number of 1-bit additions (equivalent full-adders),
   number and type of multipliers, and SRAM requirements.
The expressions in Table \ref{table:ResultsResources} 
    are based on the derived running times and resources 
    of section \ref{sec:methods} using the bit-widths described 
    in Sec. \ref{sec:setup}.
The resources of Table \ref{table:ResultsResources}
   do not include control logic 
   (e.g., for the Finite State Machine),
   or any logic needed for I/O interfacing.
We will provide more details for FPGA and SOC implementations
   in section \ref{sec:fpga}.

\begin{figure*}[!t]
\centering
\includegraphics[width=1.0\textwidth]{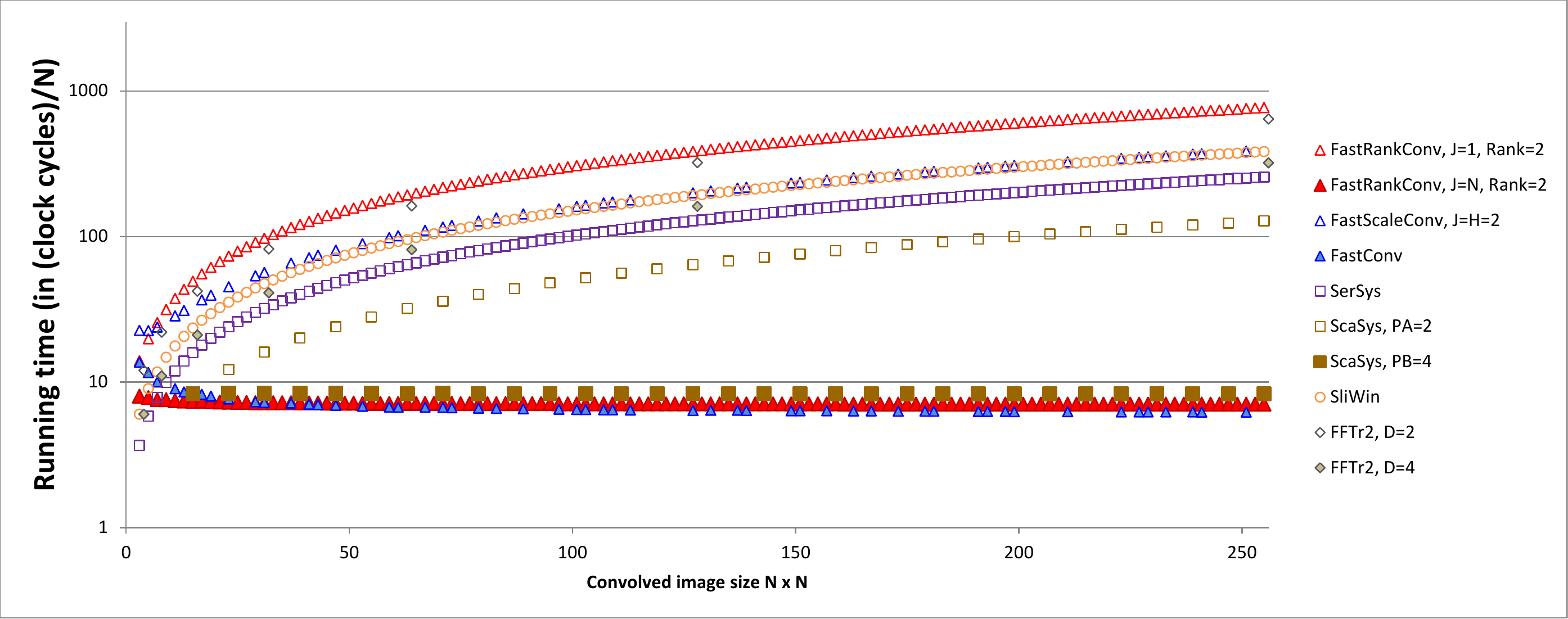}
\caption{
  Normalized running time in clock cycles versus output image (block) size ($N$).
  Here, the running time is the actual number of clock cycles divided by $N$.
  The plot refers to convolutions between $P\times P$ blocks ($N=2P-1$).
  For each implementation, as a function of $N$,
      we are assuming the use of 
      the minimum amount of required resources.
  \textit{FastConv} is the fastest followed closely by \textit{FastRankConv}
      for rank=2 and $J=N$ (approximation).
  For $J=N+1$ and $H=N$,     
      \textit{FastScaleConv} runs as fast as \textit{FastConv} and
      thus, it is not repeated here.
  Methods with $O(N)$ execution times remain below $10$ 
      (implying $10N$ execution time).
  Methods with $O(N^2)$ execution times rise well above $10$.
  Since $P=P_A \cdot P_B$,
       ScaSys implementations that achieve $O(N)$ execution time
       require $O(N^3)$ resources as opposed to $O(N^2)$ resources
       for the proposed methods (see Table \ref{table:ResultsResources}).
   Larger image sizes can also be handled using overlap-and-add.
  }
\label{fig:runtime}
\end{figure*}

\begin{figure}[!bt]
(a)\includegraphics[width=0.45\textwidth]{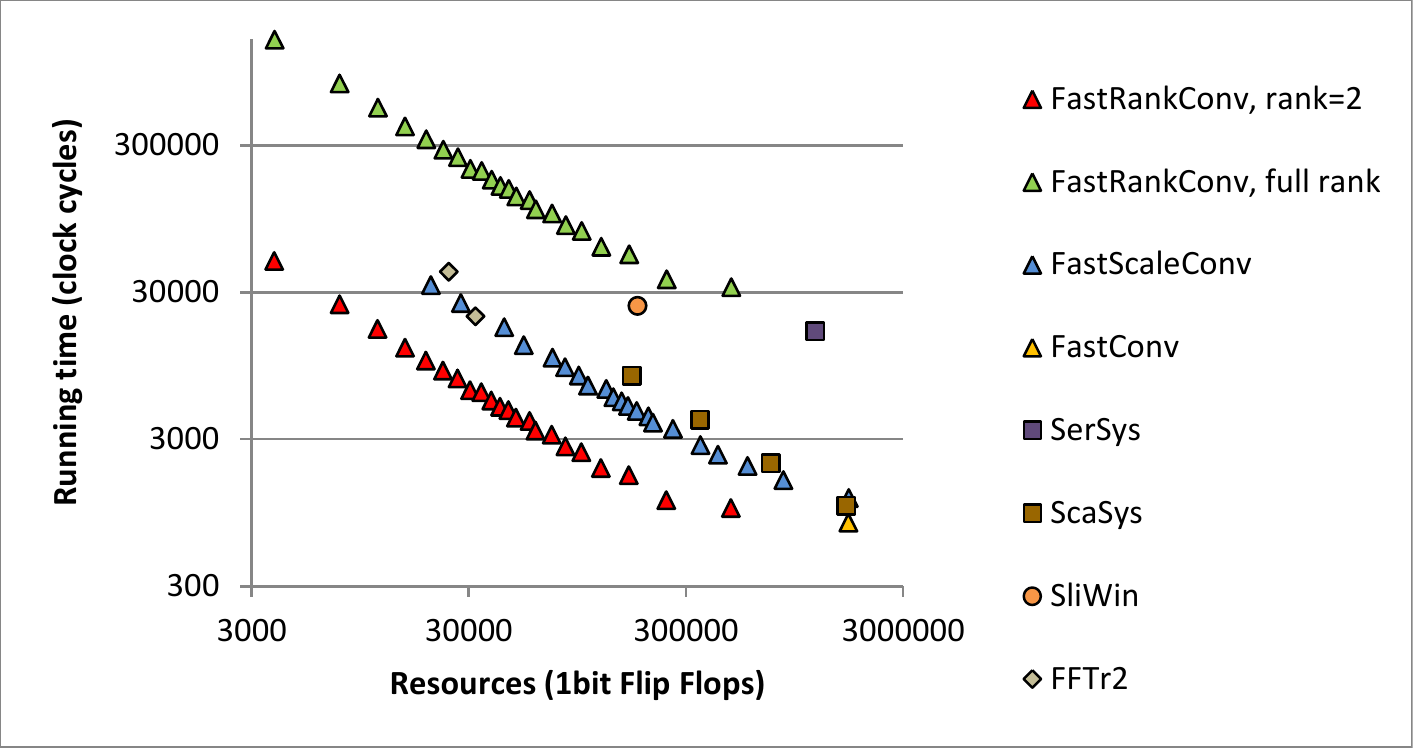}\\
~(b)\includegraphics[width=0.45\textwidth]{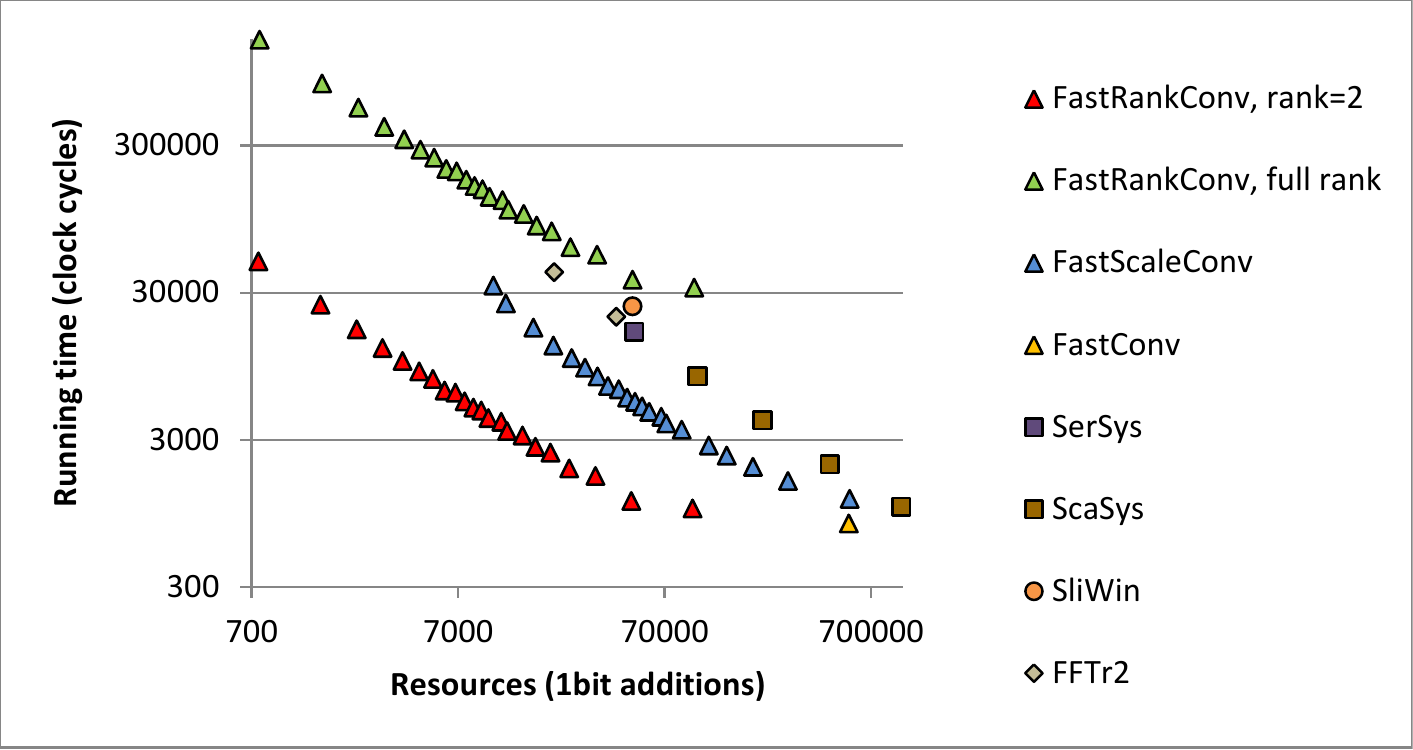}\\
~(c)\includegraphics[width=0.45\textwidth]{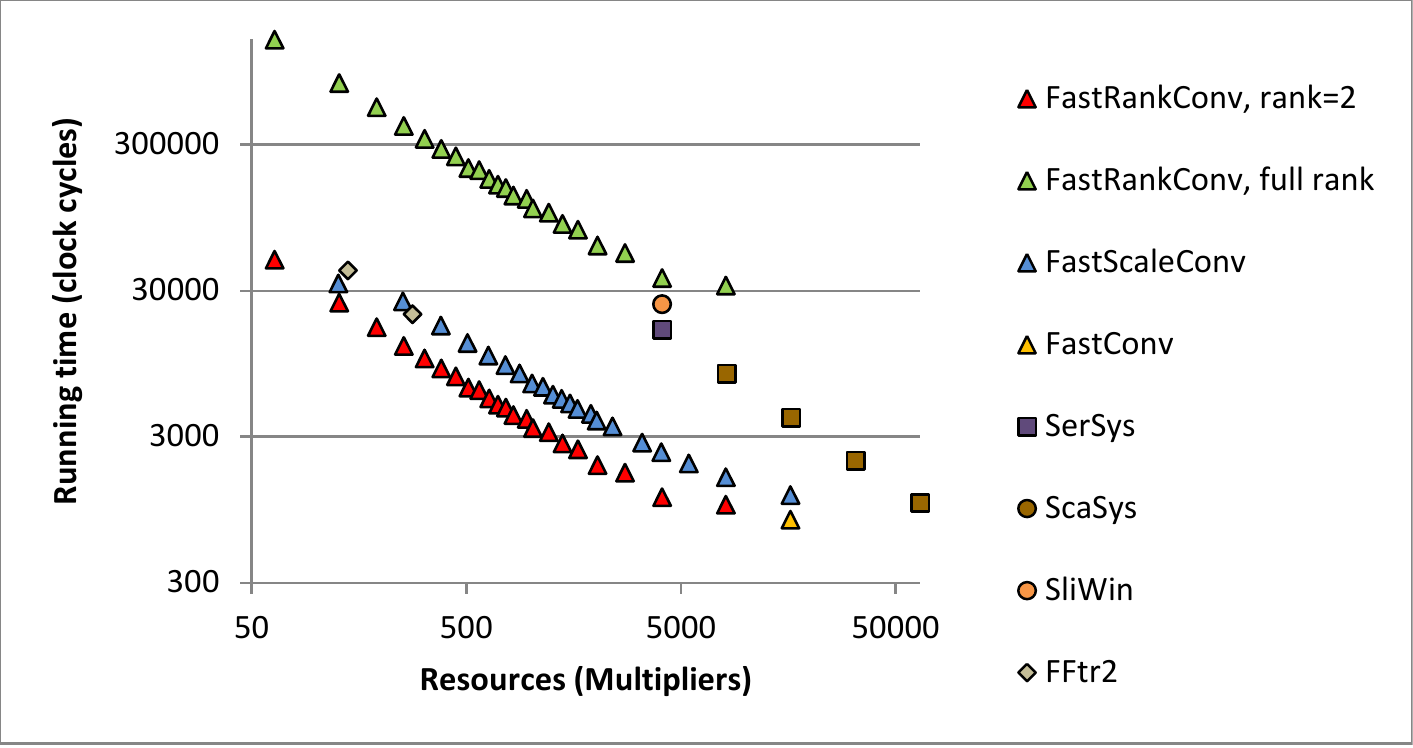}\\
\caption{
 Family of fast and scalable architectures for $N=127$ ($N=128$ for FFTr2).
 The plots refer to convolutions with $64\times 64$ blocks ($P=64$).
 For comparison purposes, we approximate
    the hardware cost for implementing 32-bit floating
    point additions and multiplications using
    fixed-point additions and 
    12-bit fixed-point multiplication
    as described in section \ref{sec:setup}.       
}
\label{fig:resultsAll}
\end{figure}

From Table \ref{table:ResultsResources}, we note
   the excellent performance of \textit{FastConv}.
When the amounts of required resources do not permit the 
   full implementation of \textit{FastConv}, 
   \textit{FastScaleConv} can be used to provide a trade-off
   between performance and required resources.
Here, note that the complex expressions for \textit{FastScaleConv}
   reduce to the ones for \textit{FastConv} when \textit{FastScaleConv}
   uses the maximum number of 1D convolvers ($J=N+1$)
   and processes the maximum number of rows in the DPRT and iDPRT ($H=N$).  
Then, as we reduce $H$ and $J$ towards 1,
   the running performance grows from $O(P)$ to $O(P^2)$ clock cycles,
   and the numbers of resources are reduced from $O(P^2)$ to $O(P)$.   
Furthermore, to interpret the rest of the table,
 from $N=2P-1$ we have that $P=(N+1)/2=P_A\cdot P_B$,
 and note that
 ${\tt A_{ffb}}(.)$, ${\tt A_{ff}}(.)$, ${\tt A_{FA}}(.)$
   grow linearly as functions of $N$.
In what follows, 
  we will further analyze performance in terms of running time (Fig. \ref{fig:runtime})
  and also provide multi-objective comparisons in terms of
  running time and required resources (Fig. \ref{fig:resultsAll}).

Before we proceed with our analysis, we note that
   the results in Figs. \ref{fig:runtime} and \ref{fig:resultsAll}
   have been derived from Table \ref{table:ResultsResources}.
On the other hand, we note that these results agree
   closely with our FPGA and SOC implementations
   that will be described in section \ref{sec:fpga}.
Furthermore, we are assuming that the data can be streamed
   to the FPGA or SOC at a rate that matches the computing rate.
Here, we note that our assumption is not very restrictive
   since the bandwidth for a PCI express 3.x    
   is about 16 GB/s and our FPGA or SOC implementations
   are running around 100MHz. 
However, for larger kernels,
   we may need custom hardware to stream
   data to the FPGA or SOC at high data rates.     

We begin with a comparison of normalized execution times
   in Fig. \ref{fig:runtime}.
For Fig. \ref{fig:runtime}, we divide
   the required number of clock cycles by $N$.
To illustrate the range of possibilities,
   we consider the two extreme performance cases;
   architectures with quadratic and linear time complexity.
    
For quadratic time complexity ($O(N^2)$ clock cycles),
   we have scalable implementations 
   derived by 
   \textit{FastScaleConv} for $J=H=2$, and
   \textit{FastRankConv} with $J=1$, $r=2$.
Alternatively, we consider a scalable extension
   of FFTr2 for $D=2, 4$, 
   a scalable implementation of ScaSys,
   and the non-scalable implementations due to SliWin and SerSys.

\textit{FastConv} provides the fastest performance
   at just $6N+5n+17$ clock cycles ($n=\log_2(N)$).
For $J=N+1$, \textit{FastScaleConv} achieves the same 
   performance as \textit{FastConv}.
For rank=2 approximations to the convolution kernel ($J=N$),     
   \textit{FastRankConv} approximates the performance of \textit{FastConv}.
In terms of related research, for $P_B=4$, 
   ScaSys achieves linear time-performance as well.
On the other hand, from Table \ref{table:ResultsResources},
   for linear time performance, using $P_A = P/P_B = P/4$, 
   we can see that ScaSys's         
   requirements grow as $P^3$ as opposed to $P^2$ growth
   for \textit{FastConv} and \textit{FastScaleConv}. 
    
Due to the significant overhead associated with
   implementing floating point arithmetic for very large
   kernels, we do not consider the extreme case where
   $N$ 1D FFTs can also yield linear performance (using
   a fast transposition, like the one we developed).
However, we do note that for very large kernels,
   as given in Table \ref{table:ResultsResources},
   the FFT based methods will most likely give the best
   results since the $N\log_2(N)$ growth in floating-point
   multipliers and additions will likely cost less
   than the $N^2$ growth of fixed-point multipliers
   and adders required by \textit{FastConv},
   \textit{FastScaleConv}, and \textit{FastRankConv}.
We have certainly not found a study in the literature
   that demonstrated that such an approach was feasible.
As we shall show next in our multi-objective comparisons,
   \textit{FastConv}, \textit{FastScaleConv}, and 
   \textit{FastRankConv} perform better than FFTr2
   in realistic convolution kernels (e.g., for $N=127$ and thus for lower $N$ also).
Furthermore, as we show later in this section,
   we can fit fast implementations of
   \textit{FastConv}, \textit{FastScaleConv}, and 
   \textit{FastRankConv} in current FPGAs and SOCs.
     
We provide detailed multi-objective
   comparisons in Fig. \ref{fig:resultsAll} for $N=127$ ($N=128$ for FFTr2).
In Fig. \ref{fig:resultsAll} we show comparisons based for
   1-bit FlipFlops (Fig. \ref{fig:resultsAll}(a)),
   equivalent 1-bit Additions (Fig. \ref{fig:resultsAll}(b)),
   and equivalent 12-bit fixed point Multipliers (Fig. \ref{fig:resultsAll}(c)).
Refer back to Table \ref{table:ResultsResources} for memory usage.
To interpret the plots, note that each curve, (termed a Pareto front),
   represents a family of optimal implementations.
The best results come from the Pareto fronts that are located 
   in the lower-left.
Within each Pareto front, the upper left point represents
   the implementation that requires the largest number of cycles (slowest)
   with the lowest number of required resources.
Then, the lower-right point represent the implementation that  
   requires the smallest number of cycles (fastest)
   with the maximum number of required resources.
To enable more direct comparisons, we also list 
   specific numbers for some of the implementations
   in table \ref{table:ResultsMultiObj}.

Since they are the fastest,
   \textit{FastConv} implementations are always in the lowest
   right portion in each plot.
From table \ref{table:ResultsMultiObj},
   we can see that \textit{FastConv}
   only requires 25\% of the multipliers and memory,
   and 56\% of the addition resources required
   by ScaSys, while requiring only 77\% of the clock-cycles.
In terms of scalable approaches,
   the Pareto front for \textit{FastRankConv} (rank=2),
   provide the best performance with minimum resources.
The limited resources required by \textit{FastRankConv}
   are also clearly documented in table \ref{table:ResultsMultiObj}.    
However, the use of rank=2 approximations may be inaccurate.
On the other hand, the full-ranked \textit{FastRankConv} requires
   the maximum amounts of resources to deliver the same performance. 

Consistently, 
   \textit{FastScaleConv} provides the best scalable implementations
   without requiring low-rank.
As seen from table \ref{table:ResultsMultiObj}, for
   the linear case, \textit{FastScaleConv}
   is slightly more expensive in resources than \textit{FastConv}
   and substantially less expensive than ScaSys.
Overall, ScaSys ($P_B=4$) implementations achieve the speed of 
   \textit{FastScaleConv} but require significantly more
   multipliers and adders.
SerSys and SliWin require significantly more resources 
   and are also much slower than \textit{FastScaleConv}.
Returning to table \ref{table:ResultsMultiObj}, for
   the quadratic case, 
   \textit{FastScaleConv} and \textit{FastRankConv} (rank=2)
   are the fastest while requiring fewer
   adders and (equivalent) multipliers.
In terms of memory requirements, our proposed scalable approaches
   do require more memory, still growing in the order of $O(N^2)$ which
   should not be a limitation with current technologies.
On the other hand, with the exception of SliWin, only
   \textit{FastConv}, \textit{FastScaleConv}
   allow the kernel to change in running time.
As a result, \textit{FastConv} and \textit{FastScaleConv}
   can also be used in cross-correlations with adaptive kernels,
   and adaptive filterbank applications.

\begin{table*}
\newcommand{\md}[2]{\multirow{#1}{0.12\textwidth}{#2}}
\newcommand{\clk}[2]{\multirow{#1}{  0.195\textwidth}{#2}}
\newcommand{\adds}[2]{\multirow{#1}{ 0.155\textwidth}{#2}}
\newcommand{\ff}[2]{\multirow{#1}{   0.22\textwidth}{#2}}
\newcommand{\mults}[2]{\multirow{#1}{0.10\textwidth}{#2}}
\newcommand{\mem}[2]{\multirow{#1}{  0.08\textwidth}{#2}}
\newcommand{\newSPC}{   \\[0.03 true in] }
\newcommand{\oldSPC}{   \\[0.05 true in] }
\caption{\label{table:ResultsResources}
Comparison of the performance 
  of 2D convolution and cross-correlations architectures
  as a function
  of computational resources.  
The result is of size of $N \times N$, where $N=2P-1$, 
  $P$ represents
  the input image size and convolution kernel size,
     $n=\left\lceil \log_2 N\right\rceil$,  
     $p=\left\lceil \log_2 P\right\rceil$,
     $J$ denotes the number of parallel 1D circular convolutions, and
     $H$ denotes the number of image rows that are processed in parallel by the DPRT.
For ScaSys, $P$ needs to be a composite number and it is assumed to be given by 
     $P=P_A\cdot P_B$.
For FFTr2, $D = 2, 4$ represents the number of 1D FFT units running in parallel.
We define:
    (i) ${\tt A_{ffb}}\left(a,b\right)$ to be number of required flip-flops inside 
          the $a$-operand of $b$ bits adder tree including input buffers,
    (ii) ${\tt A_{ff}}\left(\right)$ to be the same number without accounting for
          input buffers, and
   (iii) ${\tt A_{FA}}\left(\right)$ to be the equivalent number of 1-bit additions.
To interpret the table, note that    
   ${\tt A_{ffb}}(.)$, ${\tt A_{ff}}(.)$, and ${\tt A_{FA}}(.)$
   grow linearly as a functions of $N$,
   and can be computed exactly using the the algorithm given in
     the appendix (Fig. \ref{alg:adderResources}).
Instead of 12-bits, for the number of resources for a different number of output bits,
   simply replace 12 by the desired number of bits.
}
\begin{tabular}{lllllll}
\toprule
   \md{1}{Method} 
 & \clk{1}{Clock Cycles}  
 & \ff{1}{Flip-flops (regs for FFTr2)} 
 & \adds{1}{Additions} 
 & \mults{1}{Multipliers} 
 & \mem{1}{Memory} \\
\midrule
   \md{4}{\textit{FastConv}\\
            ~Proposed\\
            ~Fast Conv.\\
            ~(fixed point)}
 & \clk{4}{$6N+5n+17 $\\~\\~\\~ }
 & \ff{4}{$(N+1)(36N+{\tt A_{ffb}}(N,12))$ \\
            $+N(8N+{\tt A_{ff}}(N,8))$ \\
            $+12 N^2 + (N+1)\cdot{\tt A_{ff}}(N,12)$\\
            $+N(12+n)$ 
             }
 & \adds{4}{$(N+1)\cdot{\tt A_{FA}}(N,12)$\\
            $+N\cdot{\tt A_{FA}}(N,8)$\\
            $+(N+1){\tt A_{FA}}(N,12)$\\  
            $+N(12+n)$ 1-bit adds
            } 
 & \mults{4}{$(N+1) N$ 12-bit fixed point mults\\~}
 & \mem{4}{Ker: $12N(N+1)$\\~\\~\\~ }
 ~\\~\\~\\~\oldSPC
 ~\\
   \textit{FastXCorr}
 & \multicolumn{5}{c}{ 
     Same as for \textit{FastConv}.
     \textit{FastXCorr} 
     flips kernel
      prior to DPRT computations.} 
     ~\oldSPC  
   \md{7}{\textit{FastScaleConv}\\
              ~Proposed\\
              ~Scalable\\
              ~Convolution  \\
              ~(fixed point)\\
                           ~\\
                            ~}
 & \clk{7}{$\left\lceil N/H\right\rceil(N+3H+3)$\\
              $+N+\left\lceil \log_{2}H\right\rceil+1$\\
              $+\left\lceil (N+1)/J\right\rceil\cdot (J+N)$\\
                          $+n+1$ \\
                          $+\left\lceil N/H\right\rceil(N+H)$\\
                $+2\left\lceil \log_{2}N\right\rceil + \left\lceil \log_{2}H\right\rceil$\\
                $+12+3$ }
 & \ff{7}{$J\cdot (36N+{\tt A_{ffb}}(N,12))$\\
          $+N(8H+{\tt A_{ff}}(H,8))$\\  
          $+12 N(H+3)$\\ 
          $+ (N+1)\cdot{\tt A_{ff}}(H,12)$\\~\\~\\~
          }
 & \adds{7}{$J\cdot{\tt A_{FA}}(N,12)$ \\
            $+N\cdot {\tt A_{FA}}(H,8)+12N$ \\
            $+(N+1)\cdot{\tt A_{FA}}(H,12)$ \\
            $+ 2N(12+n)$\\ 1-bit adds \\~\\~
            } 
 & \mults{7}{$J\cdot N$\\ 12-bit fixed point mults\\~\\~\\~\\~\\~ }
 & \mem{7}{$24N(N+1)$ \\
           Ker: $12N(N+1)$\\~\\~\\~\\~\\~}
 \\~\\~\\~\\~\\~\\~\oldSPC
    \textit{FastScaleXCorr}
 & \multicolumn{5}{c}{ 
     Same as for \textit{FastScaleConv}.
     \textit{FastScaleXCorr} 
     flips kernel
     prior to DPRT computations.} 
     ~\oldSPC  
   \md{4}{\textit{FastRankConv}\\
          ~Proposed\\ 
          ~Fast SVD-LU\\
          ~(fixed point)}  
 & \clk{4}{$r \cdot(J+N)(\left\lceil P/J\right\rceil $ \\
           $+ \left\lceil N/J\right\rceil)+p+1$\\~\\ }    
 & \ff{4}{$J\cdot (36P+{\tt A_{ffb}}(P,12))$\\~\\~\\ } 
 & \adds{4}{$J\cdot ({\tt A_{FA}}(P,12)+12)$\\ 1-bit adds\\~\\    }
 & \mults{4}{$J\cdot P$\\ 12-bit fixed point mults\\~ }
 & \mem{4}{$8P^2+12N(N+P)$ \\ Ker: $24P^2$\\~\\~ }
 \\~\\ ~\\ ~\newSPC
 \textit{FastRankXCorr}
 & \multicolumn{5}{c}{ 
     Same as for \textit{FastRankConv}.
     Kernel flipping implemented
      during pre-processing.} 
     ~\oldSPC  
   \md{3}{SerSys \cite{Hon1990}\\~(fixed point)\\~ }
 & \clk{3}{$N^2+2P-2$\\~\\~ }
 & \ff{3}{$4P^3+34P^2-10P-12$\\~\\~ } 
 & \adds{3}{$12P(P+1)$ 1-bit adds\\~\\~ }
 & \mults{3}{$P^2$ 12-bit fixed point mults\\}
 & \mem{3}{Ker: $12P^2$\\~\\~ }
 \\ ~\\ ~\oldSPC 
   \md{3}{ScaSys \cite{Mohanty1996}\\(fixed point)\\~ }
 & \clk{3}{$\lceil N^2/P_A\rceil$\\
           $+2P_A + P_B$\\
           $+ \lceil \log_2(P \cdot P_A)\rceil$}
 & \ff{3}{$P_A(20P^2+{\tt A_{ffb}}(P_AP,12))$\\
          $+8P(P_A^2+P_A-1)$\\~ }
 & \adds{3}{$P_A(12\, P^2+$\\
            ~~~~~${\tt A_{FA}}(P_AP,12))$\\ 1-bit adds\\~ }
 & \mults{3}{$P_A\cdot P^2$ 12-bit fixed point mults }
 & \mem{3}{Ker: $12\, P_A\cdot P^2$\\~\\~ }
 \\~\\~\oldSPC
   \md{3}{SliWin \cite{Cooke2015}\\(fixed point)\\~} 
 & \clk{3}{$N \cdot P + N^2$\\
           $+ 2\left\lceil \log_2 P\right\rceil + 1$\\~ }
 & \ff{3}{$20P^2+{\tt A_{ffb}}(P^2,12)$\\~\\~ }
 & \adds{3}{${\tt A_{FA}}(P^2,12)$ 1-bit adds\\~\\~}
 & \mults{3}{$P^2$\\ 12-bit fixed point mults }
 & \mem{3}{$8 P N+8P^2+12N^2$\\~\\~ } 
 \\~\\ ~\oldSPC 
   \md{4}{FFTr2 \cite{Ayinala2012}\\
          (32-bit floating point)\\~\\~ }
 & \clk{4}{$(5N^2+4N)/D$\\~\\~\\~ }
 & \ff{4}{$D=2:$\\ 
          ~$(6N-8)$ 32-bit registers.\\
          $D=4:$\\
          ~$(8N-16)$ 32-bit registers.}
 & \adds{4}{$40D \cdot(\log_2 N + 1)$
            32-bit floating point adders.\\~}
 & \mults{4}{$2D \cdot (1$\\ $+\log_2 N)$
            32-bit float. point mults}
 & \mem{4}{$64N^2$\\Ker: $32N^2$\\~\\~}
 \\~\\~\\ ~\oldSPC      
 \bottomrule
\end{tabular}
\end{table*}

\begin{table*}
\newcommand{\md}[2]{\multirow{#1}{0.12\textwidth}{#2}}
\newcommand{\clk}[2]{\multirow{#1}{  0.195\textwidth}{#2}}
\newcommand{\adds}[2]{\multirow{#1}{ 0.155\textwidth}{#2}}
\newcommand{\ff}[2]{\multirow{#1}{   0.15\textwidth}{#2}}
\newcommand{\mults}[2]{\multirow{#1}{0.12\textwidth}{#2}}
\newcommand{\mem}[2]{\multirow{#1}{  0.15\textwidth}{#2}}
\newcommand{\newSPC}{   \\ }
\newcommand{\tblSPC}{   \\[0.03 true in] }
\caption{\label{table:ResultsMultiObj}
 Performance and resource comparisons for $N=127$ (128 for FFTr2).
 Here, we have convolutions between $64\times 64$ blocks.
 For linear-time implementations, 
     \textit{FastConv} is the fastest and serves as the reference
     design (assigned $1\times$).
 The remaining implementations are normalized by 
      the corresponding resources required by \textit{FastConv}.
 Similarly, for quadratic-time implementations,
      \textit{FastScaleConv} is used as the reference design.
Memory requirements refer to SRAM bits.
Also, note that the reported FFTr2 resources for additions
      and multiplications refer
      to an approximation of the equivalent fixed-point resources
      (refer to section \ref{sec:setup}).
}
\begin{tabular}{lllllll}
  \multicolumn{6}{l}{Implementations with linear running time, $J=128$, $H=127$,
       rank=2   (\textit{FastRankConv}), 
       $P_A=16$ (ScaSys).}\\
   \md{1}{~\textbf{Method}}     
 & \clk{1}{\textbf{Clock Cycles}}
 & \ff{1}{\textbf{Flip-flops}}
 & \adds{1}{\textbf{1-bit Additions}}
 & \mults{1}{\textbf{Multipliers}}
 & \mem{1}{\textbf{Memory}}
 \\
 \hdashline
   \md{1}{~\textit{FastConv}}
 & \clk{1}{   810     ($1\times$)}
 & \ff{1}{    1687442 ($1\times$)}
 & \adds{1}{  548101  ($1\times$)}
 & \mults{1}{ 16256   ($1\times$)}
 & \mem{1}{   195072  ($1\times$)}
 \newSPC
   \md{1}{~\textit{FastRankConv} }  
 & \clk{1}{   1023    ($1.26\times$)} 
 & \ff{1}{    484632  ($0.29\times$)} 
 & \adds{1}{  96012   ($0.18\times$)} 
 & \mults{1}{ 8128    ($0.50\times$)} 
 & \mem{1}{   422156  ($2.16\times$)} 
 \newSPC
  \md{1}{~\textit{FastScaleConv}}
 & \clk{1}{   1195    ($1.48\times$)} 
 & \ff{1}{    1689601 ($1.00\times$)} 
 & \adds{1}{  552038  ($1.01\times$)} 
 & \mults{1}{ 16256   ($1.00\times$)} 
 & \mem{1}{   585216  ($1.39\times$)} 
 \newSPC
   \md{1}{   ScaSys     \cite{Mohanty1996}}  
 & \clk{1}{   1054    ($1.30\times$)} 
 & \ff{1}{    1645888 ($0.98\times$)} 
 & \adds{1}{  982848  ($1.79\times$)} 
 & \mults{1}{ 65536   ($4.03\times$)} 
 & \mem{1}{   786432  ($4.03\times$)} 
 \\[0.1 true in]
  \multicolumn{6}{l}{Implementations with quadratic running time, 
       $J=H=4$, 
       rank=2 (\textit{FastRankConv}), 
       $D=4$  (FFTr2).}\\
   \md{1}{~\textbf{Method}}     
 & \clk{1}{\textbf{Clock Cycles}}
 & \ff{1}{\textbf{Flip-flops}}
 & \adds{1}{\textbf{1-bit Additions}}
 & \mults{1}{\textbf{Multipliers}}
 & \mem{1}{\textbf{Memory}}
 \\
 \hdashline
   \md{1}{~\textit{FastScaleConv}}
 & \clk{1}{   13093  ($1\times$)}
 & \ff{1}{    53888  ($1\times$)}
 & \adds{1}{  20309  ($1\times$)}
 & \mults{1}{ 508    ($1\times$)}
 & \mem{1}{   585216 ($1\times$)}
 \newSPC
   \md{1}{~\textit{FastRankConv}}  
 & \clk{1}{   12583 ($0.96\times$)} 
 & \ff{1}{    15264 ($0.28\times$)} 
 & \adds{1}{  3024  ($0.15\times$)} 
 & \mults{1}{ 256   ($0.50\times$)} 
 & \mem{1}{   422156 ($0.72\times$)} 
 \newSPC
   \md{1}{   SerSys     \cite{Hon1990} }  
 & \clk{1}{   16255 ($1.24\times$) }
 & \ff{1}{    1187188 ($22\times$) }
 & \adds{1}{  49908 ($2.46\times$) }
 & \mults{1}{ 4096 ($8.07\times$) }
 & \mem{1}{   49152 ($0.08\times$) }
 \newSPC
   \md{1}{   FFTr2 \cite{Ayinala2012} }  
 & \clk{1}{   20608 ($1.57\times$) }
 & \ff{1}{    33256 ($0.62\times$) }
 & \adds{1}{  40960 ($2.02\times$) }
 & \mults{1}{ 282     ($0.56\times$) }
 & \mem{1}{   1572864 ($2.69\times$) }
 \newSPC
   \md{1}{   SliWin     \cite{Cooke2015} }  
 & \clk{1}{   24270 ($1.85\times$) }
 & \ff{1}{    180212 ($3.34\times$) }
 & \adds{1}{  49140 ($2.42\times$) }
 & \mults{1}{ 4096 ($8.06\times$) }
 & \mem{1}{   291340 ($0.50\times$) }
 \newSPC
\end{tabular}
\end{table*}

\subsection{Full-precision FPGA and SOC implementations}\label{sec:fpga}
In order to understand what can be fitted in modern devices,
   we consider full-precision implementations
   for 8-bit inputs and 12-bit kernels using the on-chip memory.
Here, we assume that the image and convolution blocks are stored in the BRAMs inside the FPGA or SOC.
For overlap-and-add implementations, we assume that the images will be stored
    in external memory and be transferred to the FPGA or SOC for processing.
Here, we are not including any additional delays due to transferring the image
    from external memory to the FPGA or SOC.    
The proposed systems were 
   implemented using current FPGA and SOC technologies (Virtex-7 and Zynq-SOC).
For \textit{FastScaleConv} and \textit{FastConv}, 
   for different $N$ and $J$ (the number of parallel 1D convolvers),
   we show different implementations
   in table \ref{table:ResultsFastScaleConv}.
For \textit{FastRankConv}, by varying $P$ and $J$,
   we present different implementations in
   table \ref{table:ResultsFastRankConv}.

A collection of \textit{FastScaleConv} architectures were
   successfully implemented for $N=7$ to $N=127$.   
For $N=41$, a high-level of parallelism was achieved by
   computing the DPRT and inverse DPRT by parallel-processing
   $H=32$ rows at a time through 
   $J=32$ 1D full-precision, pipelined convolvers
   also operating in parallel.
In our full-precision example, the output images required 
   $34$ bits.
For $N=37$, we have a full precision implementation
   of \textit{FastConv} that only requires $291$ clock cycles
   by parallel processing $38$ rows of the DPRT and inverse DPRT,
   and parallel computing $38$ 1D convolutions. 
From table \ref{table:ResultsFastScaleConv}, we can see 
   that implementations are limited by the number of available
   look-up tables.
Thus, it is clear that larger values of $N$ can be implemented
   by reducing the precision requirements.
  
As shown in table \ref{table:ResultsFastRankConv},
   \textit{FastRankConv} makes a very efficient use of the DSPs
   while not requiring significant LUT resources.
For example, for $P=67$, 
   \textit{FastRankConv} only requires 16205 LUTs (out of 712000 LUTs).
In comparison, \textit{FastScaleConv}
   requirements for $N=127$ (which approximates $2P-1$),
   requires about 20 times more LUTs to deliver
   the full-accuracy results.
Also for $P=67$,  
   \textit{FastRankConv} with rank $r=2$ 
   requires $48903$ clock cycles,
   compared to $33507$ clock cycles
   for
   \textit{FastScaleConv} with $J=1$ and $H=2$
   without any rank restrictions. 
Thus, as seen earlier, for low-rank kernels,
   \textit{FastRankConv} is a good alternative 
   to \textit{FastScaleConv}.
For higher ranks and general-purpose implementations,
   \textit{FastScaleConv} is more preferable.

We also compare the results of the actual FPGA and SOC implementations of Tables
   \ref{table:ResultsFastScaleConv},
   \ref{table:ResultsFastRankConv}, and
   \ref{table:Freq}
   with the predicted measurements of Table \ref{table:ResultsResources}.
In terms of clock cycles, there is very little difference.
Here, we note that, as described in section 
   \ref{sec:methods},
   the number of clock cycles
   includes the number of cycles needed to load each image block row-by-row,
   processing, and final delay until the output is stored in the output memory. 
In terms of resources, we have found exact matches for     
   multipliers (mapped to DSPs units), flip-flops and SRAM (mapped into BRAMs).
In terms of the adders, we do have some additional combinational logic that
   is captured in the number of the LUTs.
The additional logic is due to the use of muxes in the DPRT as reported in \cite{carranza2015}.
We note that the additional overhead due to the finite state machine (FSM) and ancillary logic
   did not exceed $1\%$ of the total resources.        

We provide running times for different implementations
        in Table \ref{table:Freq}. 
  All of our implementations achieve a maximum frequency around 100MHz.
  For larger $N$, we have a slight decrease in frequency for the implementation
      of \textit{FastScaleConv}.
  The decrease is due to an increase in propagation time
      for the larger muxes used in the DPRT blocks 
      (see \cite{carranza2015}). 
  For $N=37$ and $J=38$, we have the highest
      frequency for \textit{FastConv}. 
 The highest frequency is due to the fact
      that the Fast DPRT component of \textit{FastConv} is
      a simplified version of the scalable DPRT component of
      \textit{FastScaleConv} \cite{carranza2015}.
The clock frequencies for \textit{FastRankConv}
      remain fairly constant.

\begin{figure}[!t]
\centering
\includegraphics[width=0.4\textwidth]{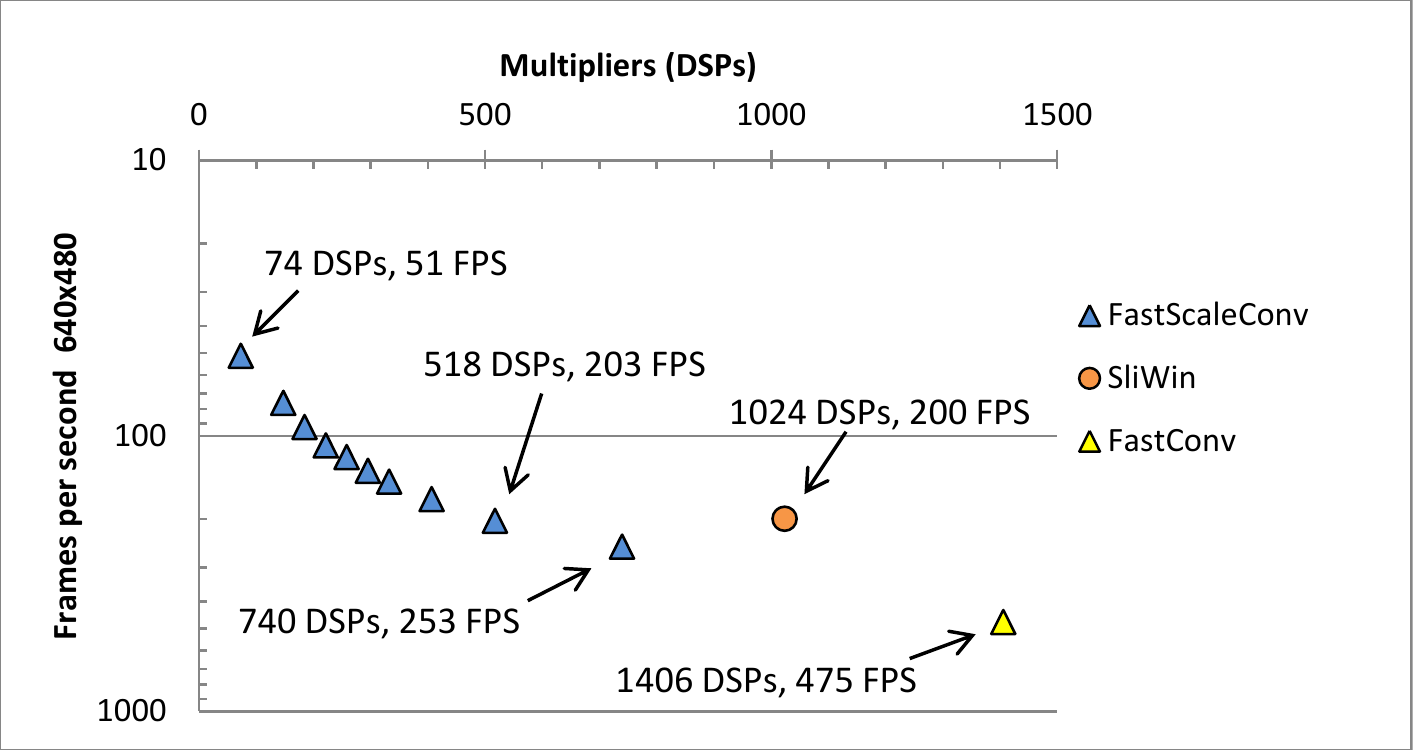}
\caption{
Performance comparison between \textit{SliWin} \cite{Cooke2015}, \textit{FastConv} and \textit{FastScaleConv}.
In terms of resources, we only count the number of DSPs which serves as
   a limiting factor for fitting the implementation in modern devices.
To measure performance, we consider the number of Frames Per Second (FPS) 
   to perform the convolution between an image of $480p$ ($640 \times 480$) 
   and a kernel of size $19 \times 19$. 
\textit{SliWin} used a Stratix IV E530 with up to $1024$ DSPs.  
\textit{FastConv} and \textit{FastScaleConv} used a Virtex-7 XC7VX1140, with up to $3360$ DSPs
  and overlap-and-add with different block sizes.
We note that it is fair to assume that both DSPs are equivalent, 
   and expect that their amounts will not change with bit-widths for inputs up to 18 bits.
At 200 FPS, \textit{FastScaleConv} uses approximately $50\%$ less DSPs than \textit{SliWin}.
On the other hand, \textit{FastConv} is 2.4 times faster with just $40\%$ more multipliers.
}
\label{fig:SliWinvsDPRT}
\end{figure}

We also provide comparisons against implementations by alternative methods.
To provide better comparisons, we compare the performance
   as a function of the number of DSPs that are required.
Based on \cite{Cooke2015}, we provide an optimized implementation for 
     \textit{SliWin}.
The comparison is given in Fig. \ref{fig:SliWinvsDPRT} for 
     images of size $640 \times 480$.
From Fig. \ref{fig:SliWinvsDPRT}, we see that 
     \textit{FastConv} remains the fastest by far
     ($2.3\times$ the \textit{SliWin} example).
Furthermore, it is clear that 
     \textit{FastScaleConv} provides a nice Pareto front
     with several optimal implementations as a function
     of the number of available DSPs.
Similar to \textit{SliWin},     
     \textit{FastScaleConv} with $H=13$, $J=14$
     achieves around $200$ FPS.
However, for this implementation,
      \textit{FastScaleConv}      
      uses approximately $50\%$ less DSPs.

\begin{table}
\newcommand{\SPC}{    \\ }
\newcommand{\tblSPC}{ \\[0.05 true in] }
\caption{\label{table:ResultsFastScaleConv}
  Full-precision implementations of 
 \textit{FastScaleConv} and
 \textit{FastConv}
 on Zynq-SoC and Virtex-7.
 Each BRAM represents up to 36 Kbits of SRAM.
 Each DSP  represents a multiplier.
 Clks refers to the required number of clock cycles
     and Bits refers to the number of bits for representing the final
     result
     for 8-bit inputs and 12-bit kernels.
 Here, $\text{Bits}=B+C+n$
    where $B=C=8$ bits and
    $n=\left\lceil \log_{2}N\right\rceil$.  
 We present a \textit{FastConv} implementation for
    $N=37$ on the Virtex-7.
 In terms of resources,
    for the Zynq-SoC (XC7Z100),
    we have 277400 LUTs, 755 BRAMs (36 Kb) and 2020 DSPs.
For the Virtex-7 (XC7VX1140),
    we have 712000 LUTs, 1880 BRAMs (36 Kb), and 3360 DSPs.   
}
\begin{tabular}{llllllll}
\toprule
  \multicolumn{8}{l}{\textbf{Zynq-SoC}}\\
  \textbf{~~~N}  &
  \textbf{Bits}  &
  \textbf{LUTs}  &
  \textbf{BRAMs} &
  \textbf{DSPs}  &
  \textbf{Clks}  &
  \textbf{J}     &
  \textbf{H} \\
  \midrule
%
 ~~~7   & 25 & 4209          & 21  & 7   & 212   & 1 & 2 \SPC
 ~~~17  & 31 & 15166         & 68  & 17  & 799   & 1 & 2 \SPC
 ~~~41  & 34 & 47271         & 205 & 41  & 3817  & 1 & 2 \SPC
 ~~~41  & 34 & 225124        & 205 & 328 & 1094  & 8 & 8 \SPC
 ~~~97  & 37 & 202925        & 485 & 97  & 19811 & 1 & 2 \SPC
 ~~~109 & 37 & 239084        & 545 & 109 & 24869 & 1 & 2 \SPC
 ~~~113 & 37 & 241106        & 565 & 113 & 26683 & 1 & 2 \tblSPC 
  \multicolumn{7}{l}{\textbf{Virtex-7}}\\
  \textbf{~~~N}  &
  \textbf{Bits}  &
  \textbf{LUTs}  &
  \textbf{BRAMs} &
  \textbf{DSPs}  &
  \textbf{Clks}  &
  \textbf{J}     &
  \textbf{H}    \\
  \midrule
%
 ~~~7   & 25 & 4125            & 21  & 7         & 212   & 1    & 2   \SPC 
 ~~~17  & 31 & 14950           & 68  & 17        & 799   & 1    & 2   \SPC
 ~~~\textbf{\textit{37}}     & 
    \textbf{\textit{34}}     & 
    \textbf{\textit{662352}} & 
    \textbf{\textit{185}}    & 
    \textbf{\textit{1406}}     & 
    \textbf{\textit{291}}   & 
    \textbf{\textit{38}}     & 
    \textbf{\textit{-}} \SPC
 ~~~41  & 34 & 46771           & 205 & 41   & 3817  & 1  & 2   \SPC
 ~~~41  & 34 & 617505          & 205 & 1312 & 658   & 32 & 32 \SPC
 ~~~127 & 37 & 321580          & 635 & 127  & 33536 & 1  & 2   \\
\bottomrule 
\end{tabular}
\end{table}

\begin{table}
\newcommand{\SPC}{    \\ }
\newcommand{\tblSPC}{ \\[0.03 true in] }
\caption{\label{table:ResultsFastRankConv}
Full-precision and scalable
   FPGA implementations of \textit{FastRankConv} (rank=2) on a 
   Virtex-7 (XC7VX1140).
The resources remain independent of rank since we
   are using a fixed a number of bits for all implementations.
Rank only affects execution times.
Refer to Table \ref{table:ResultsFastScaleConv} for definitions
    of BRAM, DSPs, Clks, and resources for the Virtex-7.   
}
\centering
\begin{tabular}{lllllll}
\toprule
  ~~~\textbf{P} & 
  \textbf{Bits} & 
  \textbf{LUTs} & 
  \textbf{BRAMs} & 
  \textbf{DSPs} & 
  \textbf{Clks} & 
  \textbf{J} \\
  \midrule
%
~~~7   & 31  & 1165   & 24   & 7    & 564    & 1  \SPC
~~~7   & 31  & 8859   & 24   & 49   & 124    & 7  \SPC
~~~17  & 35  & 3219   & 92   & 17   & 3406   & 1  \SPC
~~~17  & 35  & 59326  & 92   & 289  & 306    & 17 \SPC
~~~31  & 35  & 5737   & 169  & 31   & 11414  & 1  \SPC  
~~~31  & 35  & 180645 & 169  & 961  & 558    & 31 \SPC  
~~~41  & 37  & 8014   & 224  & 41   & 20015  & 1 \SPC
~~~41  & 37  & 353629 & 224  & 1681 & 739    & 41 \SPC
~~~53  & 37  & 9593   & 290  & 53   & 33503  & 1  \SPC
~~~53  & 37  & 513647 & 290  & 2809 & 955    & 53 \SPC
~~~61  & 37  & 14065  & 334  & 61   & 44415  & 1  \SPC
~~~67  & 39  & 16205  & 367  & 67   & 48903  & 1  \\
\bottomrule 
\end{tabular}
\end{table}

\begin{table}
\newcommand{\SPC}{    \\ }
\newcommand{\tblSPC}{ \\[0.03 true in] }
\caption{\label{table:Freq}
Selected frequencies (in Megahertz) and running time (in microseconds) for FPGA implementations using Virtex-7 (XC7VX1140).
 Refer to Tables \ref{table:ResultsFastScaleConv} and \ref{table:ResultsFastRankConv}
 for setup details.   
}
\centering
\begin{tabular}{lllllll}
\toprule
 \multicolumn{6}{l}{\textbf{\textit{FastConv}, \textit{FastScaleConv}}}\\
  ~~~\textbf{N} & 
  \textbf{J} & 
  \textbf{H} & 
  \textbf{f} & 
  \textbf{RT} & 
  \textbf{LUTs} &
	\textbf{DSPs}\\
  \midrule
~~~7   &  1  & 2  & 112  &   1.9 &   4125 &     7   \SPC
~~~17  &  1  & 2  & 109  &   7.3 &  14950 &    17   \SPC
~~~37  & 38  & -  & 113  &   2.6 & 662352 &  1406   \SPC  
~~~41  &  1  & 2  & 105  &  36.4 &  46771 &    41   \SPC
~~~127 &  1  & 2  &  98  & 342.2 & 321580 &   127   \\
\midrule
 \multicolumn{6}{l}{\textbf{\textit{FastRankConv}}}\\
  ~~~\textbf{P} & 
  \textbf{J} & 
  \textbf{rank} & 
  \textbf{f} & 
  \textbf{RT} & 
  \textbf{LUTs} &
	\textbf{DSPs}\\
  \midrule
~~~7  &  1  & 2  & 109  &   5.2 &   1165 &     7   \SPC
~~~31 &  1  & 2  & 108  & 105.7 &   5737 &    31   \SPC
~~~53 &  1  & 2  & 108  & 310.2 &   9593 &    53   \SPC
~~~53 &  53 & 2  & 107  &   8.9 & 513647 &  2809    \\
\bottomrule 
\end{tabular}
\end{table}

   By using the scalability of the proposed architectures, we can
      adjust $H$ and $J$ to ensure that the architecture fits into a given chip
      and also ensure that memory bandwidth requirements do not exceed what is available to us.
   The basic idea is to provide the proposed system with new convolution blocks after the required number of cycles.
   Fortunately, there is no bandwidth issue here.
   The required bandwidth is $O(J\cdot f)$ where $f$ denotes the operating frequency.
   For $J=2$ we have the slowest case and the required bandwidth that is $O(f)$.
   Based on our discussion, we use
       $J \approx P$ to require the highest bandwidth 
       of $O(P \times f)$, where $P \times P$ is the
			block size.
   For instance, for a video of size $640 \times 480$ at 30 FPS,
       for $P=19$, $J=H=2$ (minimum resource usage), 
		   $f = 110MHz$, 
		   each frame is processed in $19$ms (within the $33$ms to achieve the 30FPS),
		  and the required bandwidth is just $9.2MB/s$, 
		  which can be easily achieved.
   Thus, as for all 2D convolution methods, as the size of the block increases,
        our approach tends to be compute-bound.    
                
\section{Conclusions}\label{sec:conclusions}
The manuscript introduced fast and scalable
    architectures for computing
    2D cross-correlations and convolutions.
\textit{FastConv} architectures deliver
    the best performance by computing
    convolutions in $O(P)$ clock cycles.
The \textit{FastScaleConv} family of architectures
    allows us to implement efficient
    architectures that can be restricted
    to the architectures of different devices.    
The \textit{FastRankConv} family of architectures
    allows us to consider low-rank approximations
    that can significantly reduce the number
    of required resources.    
Overall, for the same level of performance,
    \textit{FastRankConv}  and
    \textit{FastScaleConv}
    require significantly fewer hardware resources
    than alternative approaches.

\section{Acknowledgments}
This material is based upon work supported by the National
Science Foundation under NSF AWD CNS-1422031.






\begin{figure}[h!]
\begin{algorithmic}[1]
\Procedure {{\tt Tree\_Resources\_WIB}}{$N,D$}
\State $n=\left\lceil \log_{2}N\right\rceil$
\State ${\tt A_{ffb}} = {\tt A_{FA}} = 0$
\State $a=N$
\For {$z$ = $1$ to $n$}
        \State $r = \left\langle a\right\rangle _{2}$
        \State $a = \left\lfloor a/2\right\rfloor$
        \State ${\tt A_{FA}} = {\tt A_{FA}} + a \cdot (D+z-1)$
        \State $a = a + r$
        \State ${\tt A_{ffb}} = {\tt A_{ffb}} + a \cdot (D+z)$
\EndFor
\State ${\tt A_{ffb}} = {\tt A_{ffb}} + X \cdot D$ \Comment{With Input Buffers (WIB)} \label{step:adderResourcesWIB}
\State \textbf{return} ${\tt A_{FA}}, {\tt A_{ffb}}$
\EndProcedure
\end{algorithmic}
\caption{\label{alg:adderResources}
Required tree resources as a function of the zero padded image ($N$),
    and the number of bits per pixel ($D$).
Refer to Table \ref{table:ResultsResources}
    for definitions of   ${\tt A_{ffb}},{\tt A_{FA}}$.
To compute ${\tt A_{ff}}$ for architectures that do not use input buffers,
   simply remove step \ref{step:adderResourcesWIB}
   from the algorithm.   
}
\end{figure}


\begin{table}
\renewcommand{\arraystretch}{1.5}
\caption{\label{table:resources}
Resource usage for different 1D Circular Convolutions implementations.
Here, we have two zero-padded images (or image blocks) 
  $g$ and $h$ of size 
  $N \times N$ (including zero-padded pixels), 
  $B$ and $C$ bits per pixel respectively and $n=\left\lceil \log_{2}N\right\rceil$.
For the adder tree, we define ${\tt A_{ffb}}$ to be the 
     number of required flip-flops including input buffers,
     and ${\tt A_{FA}}$ to be the number of 1-bit additions.
${\tt A_{ffb}}$ and ${\tt A_{FA}}$ grow linearly with respect to $N$
   and can be computed using the algorithm given in
    Fig. \ref{alg:adderResources}.
For the multipliers, we note that each one is
    implemented using two inputs of size $B+n$ and $C+n$ bits
    and an output of $B+C+2n$ bits.
        Here, we use the term ``1-bit additions"        to refer to the number
        of equivalent 1-bit full adders.
}
\begin{center}
\resizebox{0.5\textwidth}{!}{%
\begin{tabular}{llll}
\toprule
Block & Number of flip-flops & 1-bit additions & Multipliers \\
\midrule
 Core  &       $N(2B+2C+5n)$ & ${\tt A_{FA}}(N,B+C+2n)$ &      $N$  \\
 \hfill & $+{\tt A_{ffb}}(N,B+C+2n)$ & \hfill & \hfill  \\
 System &       $J\cdot N(2B+2C+5n)$ & $J\cdot {\tt A_{FA}}(N,B+C+2n)$ &    $J\cdot N$  \\
 \hfill & $+J{\tt A_{ffb}}(N,B+C+2n)$ & \hfill & \hfill \\
 \bottomrule
\end{tabular}}
\end{center}
\end{table}

\begin{table}
\renewcommand{\arraystretch}{1.5}
\caption{\label{table:resourcesLU}Resource usage for different Linear Convolvers implementations.
Here, all the quantities are given for maximum accuracy.
Refer to the caption of Table \ref{table:resources} and section \ref{sec:notation} for the notation.
}
\begin{center}
\resizebox{0.5\textwidth}{!}{%
\begin{tabular}{llll}
\toprule
Block & Number of flip-flops & 1-bit additions & Multipliers \\
\midrule
 Core  &       $N2 \cdot (B+C+q2)+Q2 \cdot C $ & ${\tt A_{FA}}(Q2,B+2C+q2)$ &       $Q2$  \\
 \hfill & $+{\tt A_{ffb}}(Q2,B+2C+q2)$ & \hfill & \hfill  \\
 \hline
 System        &       $J \cdot  (N2 \cdot (B+C+q2)+Q2 \cdot C $ & $J \cdot {\tt A_{FA}}(Q2,B+2C+q2)$ &      $J \cdot  Q2$  \\
 \hfill & $+{\tt A_{ffb}}(Q2,B+2C+q2))$ & \hfill & \hfill  \\
 \bottomrule
\end{tabular}}
\end{center}
\end{table}

\ifCLASSOPTIONcaptionsoff
  \newpage
\fi

\bibliographystyle{IEEEtran} 
\bibliography{convolution-v06b} 

\begin{IEEEbiography}[{\includegraphics[width=1in,height=1.25in,clip,keepaspectratio]{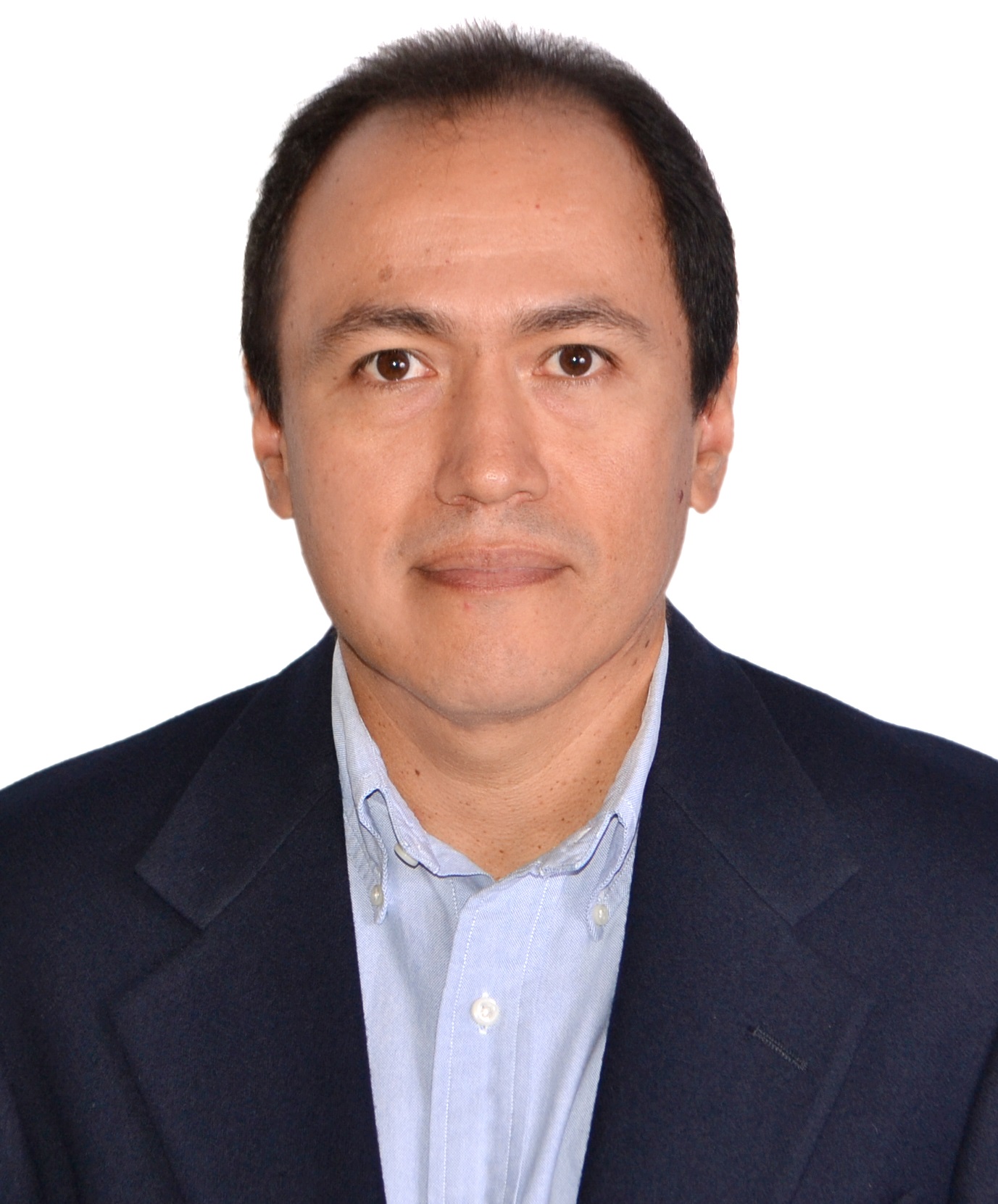}}]{Cesar Carranza}
Cesar Carranza received his 
  Ph.D. degree in Engineering (with distinction) and his M.Sc. degree in Computer Engineering 
          from the University of New Mexico at Albuquerque in 2016 and 2012 respectively.
        He also holds a M.Sc in Computer Science
          from Centro de Investigaci\'on Cient\'ifica y de Educaci\'on Superior de Ensenada in 2010
        and a B.Sc. in Electrical Engineering from Pontificia Universidad Cat\'olica del Per\'u in 1994.
He is currently an Associate Professor at Pontificia Universidad Cat\'olica del Per\'u.
His current research interests include parallel algorithms for image processing, high performance hardware integration and parallel computing.

\end{IEEEbiography}

\begin{IEEEbiography}[{\includegraphics[width=1in,height=1.25in,clip,keepaspectratio]{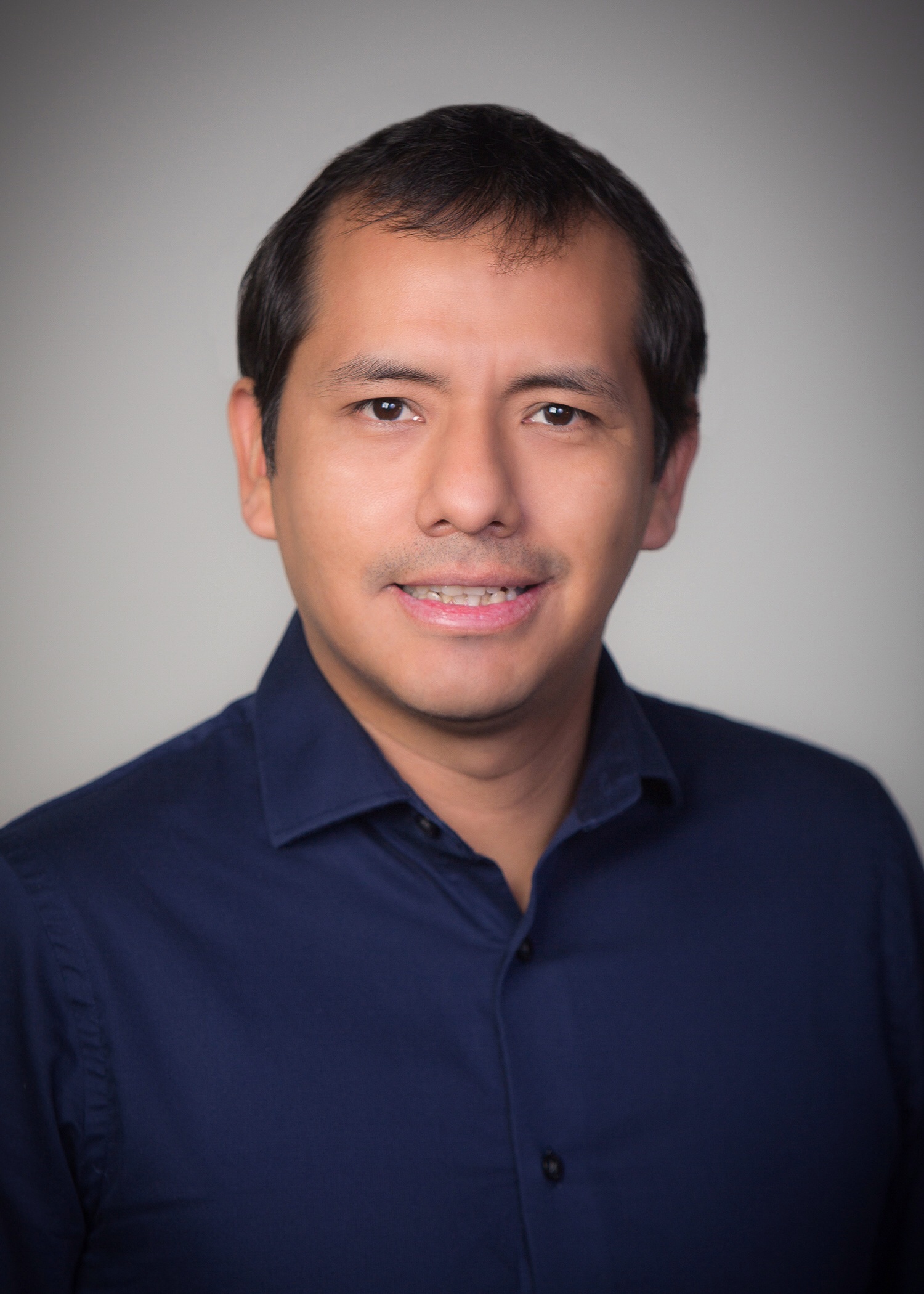}}]{Daniel Llamocca}
Daniel Llamocca received his Ph.D. degree in Computer Engineering and his M.Sc. degree in Electrical Engineering from the University of New Mexico at Albuquerque in 2012 and 2008 respectively. He also holds a B.Sc. in Electrical Engineering from the Pontifical Catholic University of Peru in 2002.

He is currently an Assistant Professor at Oakland University. His research deals with run-time automatic adaptation of hardware resources to time-varying constraints with the purpose of delivering the best hardware solution at any time. His current research interests include: i) reconfigurable computer architectures for signal, image, and video processing, ii) high-performance architectures for computer arithmetic, communication, and embedded interfaces, iii) embedded system design, and iv) Run-time Partial Reconfiguration techniques on FPGAs.
\end{IEEEbiography}

\begin{IEEEbiography}[{\includegraphics[width=1in,height=1.25in,clip,keepaspectratio]{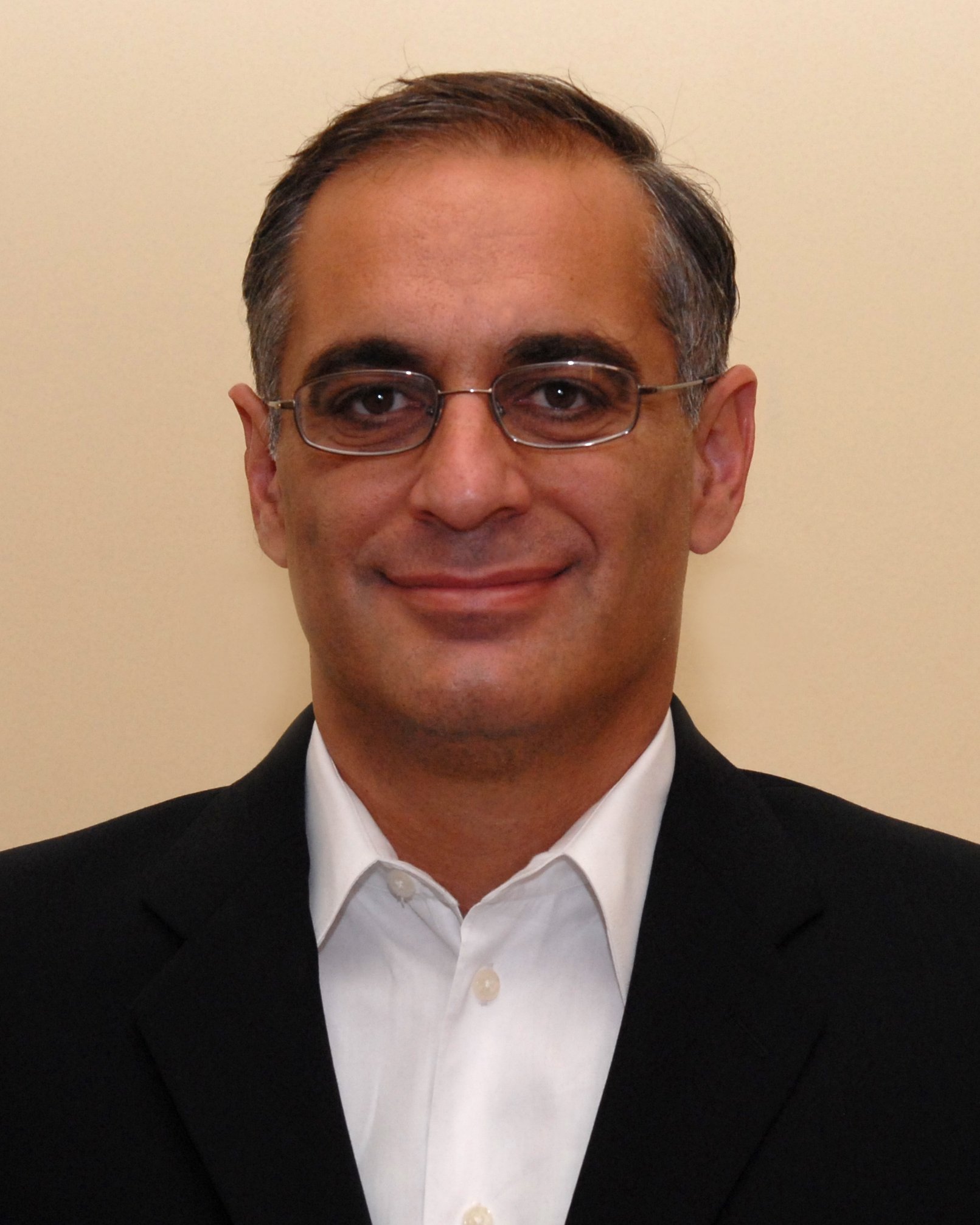}}]{Marios Pattichis}
Marios Pattichis (M'99, SM'06) received the B.Sc.(High Hons. and Special Hons.) degree in computer sciences and the B.A. (High Hons.) degree in mathematics, both in 1991, the M.S. degree in electrical engineering in 1993, and the Ph.D. degree in computer engineering in 1998, all from the University of Texas, Austin. He is currently a Professor with the Department of Electrical and Computer Engineering, University of New Mexico (UNM), Albuquerque. His current research interests include digital image, video processing, communications, dynamically reconfigurable computer architectures, and biomedical and space image-processing applications.

Dr. Pattichis is currently a senior associate editor for the
    \textit{IEEE Signal Processing Letters}.
He has been an associate editor for the 
    \textit{IEEE Transactions on Image Processing},
    \textit{IEEE Transactions on Industrial Informatics}, 
    and has also served as a guest associate editor for the 
    \textit{IEEE Transactions on Information Technology in Biomedicine}. 
He was the general chair of the 
    \textit{2008 IEEE Southwest Symposium on Image Analysis and Interpretation}. 
He was a recipient of the 
  2016 Lawton-Ellis and the 
  2004 distinguished teaching awards from the department 
  of electrical and computer engineering at UNM. 
For his development of the digital logic design labs at 
  UNM, he was recognized by the Xilinx Corporation in 2003 and by the UNM 
  School of Engineering's Harrison faculty excellence award in 2006. 
He was a founding Co-PI of the Configurable Space Microsystems Innovations
   \& Applications Center (COSMIAC) at UNM. 
At UNM, he is currently the director of the image and video Processing 
   and Communications Lab (ivPCL, \url{ivpcl.unm.edu}).
\end{IEEEbiography}

\end{document}